\documentclass[11pt]{iopart}

\usepackage{harvard}
\usepackage{url}
\usepackage{float}
\usepackage{xcolor}
\usepackage{colortbl}
\usepackage{hyperref}
\usepackage{graphicx}

 \definecolor{1}{rgb}{0.45, 0.75, 0.45} 
 \definecolor{2}{rgb}{0.7, 0.9, 0.6} 
 \definecolor{3}{rgb}{1.0, 0.8, 0.6} 
 \definecolor{4}{rgb}{1.0, 0.6, 0.6} 

\begin{document}

\title[Anatomical feature-prioritized loss for MR to CT translation]{Anatomical feature-prioritized loss for enhanced MR to CT translation}

\author{Arthur Longuefosse$^1$, 
Baudouin Denis de Senneville$^2$, 
Gaël Dournes$^3$, 
Ilyes Benlala$^3$, 
Fabien Baldacci$^1$, 
Pascal Desbarats$^1$}

\address{$^1$ Univ. Bordeaux, CNRS, Bordeaux INP, LaBRI, UMR 5800, 33400 Talence, France}
\address{$^2$ Univ. Bordeaux, CNRS, Bordeaux INP, IMB, UMR 5251, 33400 Talence, France}
\address{$^3$ Service d’Imagerie Médicale Radiologie Diagnostique et Thérapeutique, CHU de Bordeaux, 33000 Bordeaux, France}

\ead{arthur.longuefosse@gmail.com}

\vspace{10pt}

\begin{abstract}

\noindent \textit{Objective.} Accurate reconstruction of localized anatomical details is essential in medical image synthesis, particularly when addressing specific clinical requirements such as the identification or measurement of fine structures. Traditional methods for image translation and synthesis are generally optimized for global image reconstruction but often fall short in providing the finesse required for detailed local analysis. This study represents a step toward addressing this challenge by introducing a novel anatomical feature-prioritized (AFP) loss function into the synthesis process. \textit{Approach.} The AFP loss integrates features from pre-trained task-specific models, such as anatomical segmentation networks, into the image synthesis pipeline to enforce attention to critical structures. This loss function is evaluated across multiple architectures, including GAN-based and CNN-based models, and applied in two cross-modality contexts: (1) lung MR to CT translation with an emphasis on bronchial structure preservation, using a private thoracic dataset; and (2) pelvis MR to CT synthesis, targeting organ and muscle reconstruction, using the public SynthRAD2023 dataset. Feature embeddings from domain-specific segmentation networks are extracted to guide synthesis toward anatomically meaningful outputs. \textit{Main results.} The AFP loss demonstrated consistent improvements in downstream segmentation accuracy across both domains. For lung airway reconstruction, the Dice coefficient increased from 0.534 with standard L1 loss to 0.584 using AFP loss. In pelvic imaging, bone reconstruction Dice scores improved from 0.738 using L1 loss to 0.780 with AFP loss. These results confirm that the AFP loss improves the reconstruction of anatomical structures while maintaining comparable intensity-based metrics, indicating that global image quality is not compromised. \textit{Significance.} The proposed AFP loss provides a modular and generalizable approach for embedding anatomical task-awareness into medical image synthesis. By aligning image translation objectives with clinically relevant features, it offers a pathway toward more precise and useful synthetic images for downstream tasks, supporting broader integration of image synthesis in clinical workflows.

\end{abstract}

%
\vspace{2pc}
\submitto{\PMB}
%
%
%

\section{Introduction}
Over the past decade, medical image synthesis has emerged as a significant trend in the field of healthcare \cite{generative_medical}. This technique plays a pivotal role in producing synthetic data to address the shortfall in medical datasets, such as data diversity \cite{synthetic_augmentation}, and improving the robustness of machine learning models. Additionally, it serves as a valuable tool for translating between different imaging modalities, offering benefits in applications like attenuation correction or radiation therapy planning, as described in \cite{synthrad_report}. The significant advancement of this field is mainly attributed to the evolution of deep learning, notably the advent of generative methods, such as generative adversarial networks (GANs) \cite{gan}, which enable realistic image generation through adversarial training, and diffusion models (DMs) \cite{dm}, that offer greater stability by iteratively refining images from noise. Many recent studies have explored synthetic data generation \cite{synthetic_dataset} as well as cross-modality translation \cite{medgan}, addressing synthesis challenges in anatomical areas such as brain \cite{brain}, aorta \cite{aorta}, lungs \cite{longuefosse_radiology}.

Among these generative methods, Convolutional Neural Networks (CNNs), such as U-Net \cite{u-net}, have been particularly effective due to their ability to capture both local and global features efficiently. The encoder-decoder architecture with skip connections enables precise localization, making U-Net particularly effective for tasks requiring fine structural preservation. Beyond CNNs, GANs and their conditional variants (e.g., pix2pix and cycleGAN) have significantly advanced image-to-image translation by generating high-quality images. Pix2pix \cite{pix2pix} learns paired data mappings, while cycleGAN \cite{cyclegan} extends this capability to unpaired datasets, increasing its versatility. Recent approaches such as SPADE \cite{spade} have further improved semantic image synthesis by incorporating spatially-adaptive normalization, which has demonstrated effectiveness in 3D medical imaging \cite{spade3D}. More recently, diffusion models have gained prominence due to their ability to generate high-resolution images by iteratively refining image quality through a stochastic process \cite{dm}. Latent Diffusion Models (LDMs) \cite{benchmark_dm} have shown potential in cross-modality medical image synthesis \cite{cross_dm}, allowing efficient and high-quality image translation. However, diffusion models remain computationally demanding and are often limited to slice-wise processing, making their application to full volumetric medical images challenging.

Despite these advances, there remains a gap in the precision and fidelity of synthesized images, particularly in the reconstruction of detailed anatomical structures. Most current state-of-the-art methods, including those based on conditional generative models (e.g. cGANs introduced in \cite{cgan}) focus on optimizing global image reconstruction, often at the expense of accurately capturing the nuances of specific features. This is largely due to the reliance on global loss functions such as the L1 loss, which primarily measures global intensity differences and does not account for the preservation of anatomical structures and critical medical features. This kind of loss also tends to smooth out finer details, leading to images that may appear realistic at a macro level but lack the precision required for clinical tasks involving the identification and precise measurement of small, intricate structures. This oversight becomes particularly critical when addressing explicit clinical needs, such as the accurate delineation of boundaries between different tissues, the identification of lesions, or the precise measurement of anatomical structures for surgical planning or disease monitoring. 

To address the need for structure-preserving synthesis, several works have explored anatomy-aware approaches. In \cite{struct-gan}, authors use a structure-constrained GAN for unsupervised MR-to-CT synthesis by defining a structure-consistency loss based on the modality independent neighborhood descriptor, introduced in \cite{mind}, often employed for cross-modality registration. \cite{Roi-GAN} employs a multi-task network for synthetic CT generation from MRI, emphasizing bone density value prediction. It uses a composite loss function to localize regions of interest, combining classification and regression tasks to improve performance and computational efficiency, particularly in radiation therapy planning. \cite{SA-GAN} introduces a structure-aware GAN for organ-preserving synthetic CT generation from MRI, using a dual-stream approach based on a segmentation loss to maintain organ integrity and achieve clinically acceptable accuracy in MR-only treatment planning. More recently, UNest \cite{unest} introduced structural attention using Segment Anything to guide transformer-based synthesis, and HiFi-Syn \cite{hifisyn} proposed hierarchical discrimination across pixel, structure and global levels for high-fidelity MR synthesis.

These methods focus on specific body parts or anatomical structures using segmentation networks but often do not fully leverage the capabilities of these networks. Typically, segmentation networks are employed for their final output labels \cite{labelinformed}, with segmentation loss guiding training. However, this approach usually relies on the activation of the last layer of the network, which represents a coarse classification of anatomical regions. It neglects the rich, multi-layered feature representations extracted throughout the entire network during the segmentation process. These intermediate features capture essential structural information, offering detailed insights into the anatomy that are not fully utilized when relying solely on the final output. Addressing these limitations, our study introduces a novel approach aimed at enhancing the synthesis of specific medical details across different anatomical regions. We introduce an anatomical feature-prioritized loss function, denoted AFP loss, specifically designed for medical image translation, ensuring both global fidelity and detailed structural accuracy. This approach leverages a distance between features extracted from a model pre-trained on a specific medical task, such as the segmentation of a delimited region, based on the idea that such a model presents an advanced understanding and representation of those medical structures.

A natural choice for implementing AFP loss is nnU-Net, a self-configuring CNN that has demonstrated state-of-the-art performance in medical image segmentation across various datasets \cite{nnU-Net}. Recent studies have shown that nnU-Net can be adapted for synthesis tasks, often outperforming GAN-based models \cite{adapted_nnU-Net}. Unlike GANs, which rely on adversarial losses prone to instability, nnU-Net provides a robust baseline that allows full control over loss function design. This flexibility enables us to integrate AFP loss seamlessly, either as a standalone constraint or in combination with traditional losses like L1 loss, allowing us to emphasize fine structural preservation without being limited by adversarial training instability.

In this work, we validate AFP loss across different synthesis models, including both generative approaches (e.g., pix2pixHD \cite{pix2pixhd}, SPADE \cite{spade}) and CNN-based architectures (e.g., nnU-Net \cite{nnU-Net}). We evaluate our method on two distinct tasks: MR-to-CT translation for lung imaging, where preserving fine structures such as airways and vessels is critical, and MR-to-CT synthesis in the pelvis, where coarser anatomical features like muscles and organs must be maintained. Through extensive qualitative and quantitative evaluations, we demonstrate that AFP loss significantly enhances anatomical fidelity, making it a valuable contribution to medical image synthesis for clinical applications.

\section{Method}
\subsection{Anatomical Feature-Prioritized Loss Function}
\begin{figure*}[!h]
  \label{fig:layers}
  \centering
  \caption{Example of MR to CT translation pipeline demonstrating the application of the anatomical feature-prioritized loss. A conditional generative model synthesizes a CT image from the input MR. The proposed AFP loss is calculated by comparing activation maps of synthesized and real CT from a pre-trained U-Net specialized in airway segmentation. All features are acquired after the ReLU activation in each convolutional block. The AFP loss can either replace or complement traditional reconstruction losses.}
  \includegraphics[width=1\linewidth, trim = 0cm 17cm 0cm 0cm, clip]{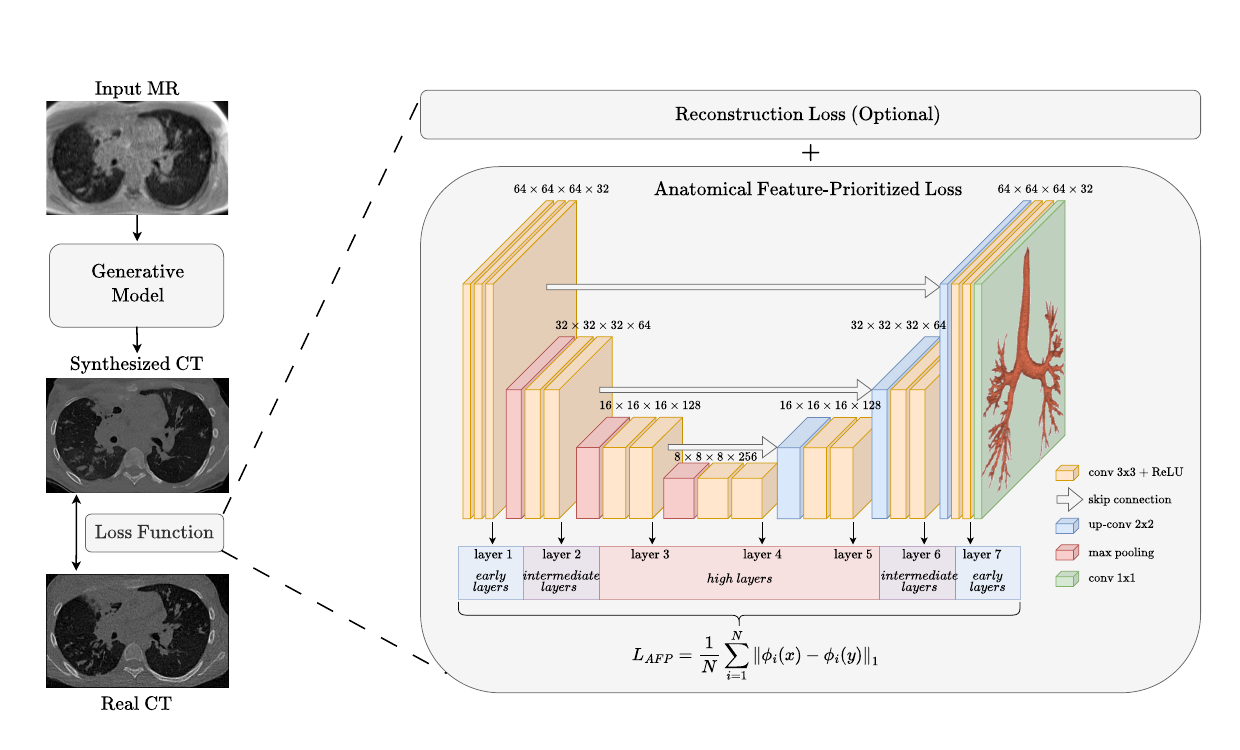}
\end{figure*}

In this study, we introduce a novel anatomical feature-prioritized (AFP) loss function. The main goal of this loss is to enhance the synthesis of specific anatomical structures in medical images, which are often inadequately reconstructed when using standard pixel-based reconstruction losses. Our approach leverages pre-trained segmentation network, which are specialized in identifying and segmenting precise anatomical regions. By extracting features from these networks, we focus not only on the overall image similarity but also on preserving medically relevant structures, such as organs, vessels, and bones. The mechanism of our feature-prioritized loss is inspired by perceptual loss \cite{perceptual}, commonly used in natural image synthesis, which focuses on minimizing the difference between the feature representations of synthetic and real images extracted from various layers of a deep network. Traditionally, perceptual loss is defined using 2D classification networks on natural images, such as VGG on ImageNet, and its use in 3D is often limited. In our case, we use feature representations from segmentation networks that are trained to understand the structures of medical images. \\
In contrast to classification networks, which are designed to categorize images into general classes (e.g., identifying whether an image contains a dog or a cat), segmentation networks are specifically trained to understand and represent detailed structural information needed for anatomical delineation. The encoder in a segmentation network extracts hierarchical features that are directly relevant for identifying and segmenting anatomical structures, whereas classification networks focus on more abstract, high-level features. Moreover, segmentation decoders allow the model to map these features into specific anatomical structures, which are compared to ground-truth segmentation labels. By using features from a segmentation network, our AFP loss prioritizes medically relevant structures, making it a more effective choice for medical image synthesis tasks. \\
The AFP loss is defined as :
\begin{equation}
\mathcal{L}_{AFP (x,y)} = \frac{1}{N} \sum_{i=1}^{N} \left\| \phi_i(x) - \phi_i(y) \right\|_1
\end{equation} 
Here, $x$ and $y$ represent the synthesized and real images to compare, $N$ denotes the number of layers extracted from the pre-trained network and $\phi_{i}$ the features of the \textit{i}-th layer. The L1 norm is applied to the feature differences, ensuring that the model minimizes the difference between the feature representations of the synthesized and real images. \\
The AFP loss is computed using L1 loss in the feature space rather than in the pixel space. By comparing the feature representations of the synthesized and real images, we ensure that the model focuses on preserving high-level anatomical features rather than just pixel-wise intensity differences. The use of L1 loss in feature space is a well-established method in perceptual loss formulations, where it has been shown to improve the alignment of synthesized images with human perception of structure and detail \cite{deep_features}. By leveraging feature representations from a segmentation network, we ensure that the AFP loss emphasizes medically relevant features, making it a more suitable choice for tasks like MR-to-CT translation, where preserving fine anatomical details is essential. Figure \ref{fig:layers} illustrates an example of an MR to CT translation pipeline, where the AFP loss is applied to a U-Net network trained for airway segmentation.

\subsection{Integration in synthesis networks}
The AFP loss introduced in this study is a versatile function, adaptable to various synthesis and translation tasks within medical imaging. Depending on the specific synthesis task, the AFP loss can be combined with a global reconstruction loss, such as the traditional L1 loss, to maintain a correct range of intensities. A general recommendation is to start with a global network trained with L1 loss to establish a baseline intensity reconstruction and then fine-tune the model with our AFP loss to refine the anatomical structures. This approach ensures a two-step process: an initial global generation step followed by a local anatomical refinement step.

For tasks involving the generation of fine structures, such as vessels or bronchi, it may be preferable to use only the AFP loss from the beginning. In such cases, L1 loss should be avoided due to its tendency to produce blurred reconstructions, which can obscure fine details. The AFP loss is designed to prioritize anatomical fidelity, aiming to enhance the preservation of fine structures while being less sensitive to global intensity variations.

While we illustrate this loss using pre-trained U-Net models, our approach can be adapted to any CNN trained on medical data. The AFP loss serves as a robust addition to medical image synthesis and translation tasks, enhancing the capability to generate high-fidelity anatomical structures.

\subsection{Implemented Synthesis Models}
To evaluate the performances of our anatomical structure prioritized loss function (AFP loss), we implemented it in two primary models for medical image synthesis: a GAN-based model and a CNN-based models. Specifically, we develop a 3D variant of the pix2pixHD model and adapt a 3D nnU-Net model for translation tasks.  

\subsubsection*{GAN-based model}
We develop a 3D variant of the pix2pixHD model \cite{pix2pixhd}, incorporating 3D versions of the generator and discriminator, as well as 3D SPADE layers \cite{spade}. This architecture facilitates image-to-image translation by incorporating spatially-adaptive normalization, effectively mitigating the issue of false positives encountered in the pix2pixHD network alone \cite{longuefosse_isbi}. For network optimization, we use the default parameters of the SPADE model: Adam optimizer with $\beta_1 = 0.5$ and $\beta_2 = 0.999$, an initial learning rate of 0.0002, and a batch size of 2. For the training process, we employ a composite loss function that includes an adversarial loss, a feature-matching loss, and a global reconstruction loss. The adversarial loss utilizes a Hinge loss mechanism, as adopted in the SPADE implementation \cite{spade}, to refine the adversarial training dynamics. The feature-matching loss ensures alignment between the discriminator's intermediate representations of the real and synthesized images, utilizing an L1 distance to promote stability and fidelity in the generated images \cite{pix2pixhd}. A global reconstruction loss is incorporated to further enhance the overall quality of the synthesized output, defined as an L1 loss. Finally, our AFP loss is incorporated to complement the L1 loss, enhancing localized structural details, or to replace it entirely, offering a more targeted approach to preserving anatomical accuracy in the synthesized images. 

\subsubsection*{CNN-based model}
We have extended the application of the nnU-Net framework \cite{nnU-Net}, traditionally a pioneer in medical image segmentation, to tackle translation tasks. Our previous work \cite{adapted_nnU-Net} provides an adaptation of the nnU-Net for synthesis tasks and showed promise with a standard MSE loss function. The adapted model retains the original framework's preprocessing and low-complexity architecture, which benefits from not being dependent on a discriminator, a component known for its training challenges. We propose a new version of the adapted nnU-Net incorporating the proposed AFP loss in the training process. This setup enables us to precisely quantify the impact of our AFP loss, whether used alone or paired with an L1 loss, providing a robust framework for improving medical image synthesis. The model has been made publicly available at \url{https://github.com/Phyrise/nnUNet_translation}.
For network optimization, we use the default nnU-Net parameters: SGD with momentum 0.99 and Nesterov acceleration, an initial learning rate of 0.01, a weight decay of 0.0001 for regularization, and a batch size of 2.

\subsection{Reconstruction from patches}
To address the memory constraints associated with processing 3D models, we adopt a patch-based approach for each synthesis model. Patch size is automatically determined by nnU-Net’s built-in optimization based on available GPU memory. To ensure smooth reconstruction and mitigate border effects between patches, we use patches with 50\% overlap in each spatial dimension. During inference, we apply median fusion on the voxel intensities in overlapping regions. This method helps preserve high-frequency details, resulting in a visually more accurate and detailed final output. While mean fusion tends to smooth the output and can yield slightly better intensity-based scores, median fusion better maintains anatomical sharpness and fine structures, which is crucial for downstream tasks.

\subsection{Artifact Reduction Strategies}
The AFP loss operates similarly to a perceptual loss, which has been shown to potentially lead to checkerboard artifacts \cite{checkerboard}, primarly due to transposed convolution operations in the decoder part. These artifacts often appear as high-frequency grid-like patterns in the reconstructed images and can compromise visual quality as well as downstream performance.
To address this issue, we explore the effectiveness of combining the AFP loss with a traditional L1 loss, which can be effective in reducing high-frequency errors that contribute to checkerboard patterns. This approach aim to leverage the global intensity preservation of the L1 loss while benefiting from the detailed structure preservation of the AFP loss. We also investigated replacing transposed convolutions with a combination of upsampling and convolutional layers as proposed by \cite{upsampling}. This proven technique helps to reduce checkerboard artifacts by providing a more stable upsampling mechanism, particularly when using feature-based losses.

To address this issue, we explore the effectiveness of combining the AFP loss with a traditional L1 loss. While AFP loss emphasizes the preservation of anatomical structure by comparing features extracted from segmentation networks, it does not explicitly address pixel-wise intensity differences. Conversely, L1 loss, being a pixel-wise intensity-based metric, promotes global intensity consistency and smoothness. By combining the two, we aim to benefit from the complementary properties of both losses: AFP ensures structural consistency in critical anatomical regions, while L1 suppresses high-frequency noise and smoothens transitions, reducing the likelihood of checkerboard patterns.

In addition, we investigated architectural changes to further mitigate these artifacts. Specifically, we replaced transposed convolution operations in the decoder with a combination of bilinear upsampling followed by standard convolution layers, as suggested by \cite{upsampling}. This approach avoids the uneven kernel overlap problem inherent to transposed convolutions, providing a more stable and artifact-free upsampling process, particularly beneficial when feature-based losses like AFP are employed.

\section{Experiments}
\subsection{Datasets}
\subsubsection*{Lung MR to CT Synthesis}
For lung MR to CT synthesis, we used a private dataset of thoracic images from UTE MR and CT scans of 122 patients. Both modalities were acquired on the same day, from 2018 to 2022. CT images were obtained using a Siemens SOMATOM Force and a GE Medical Systems Revolution CT in end-expiration, with sharp filters. The acquisition parameters were a DLP of 10 mGy.cm, 120 kV, and a tube current ranging from 10 to 475 mA. Images were reconstructed using the SAFIRE iterative reconstruction, with a voxel size of 0.6 $\times$ 0.6 $\times$ 0.6 mm$^3$.  The UTE MR images were acquired using the SpiralVibe sequence on a SIEMENS Aera scanner, with the following parameters: TR/TE/flip angle=4.1ms/0.07ms/5°, with a voxel size of $1\times 1\times 1$ mm$^3$. The acquisition matrix ranged from $320-352 \times 320-352 \times 180-260$ and the bandwith was set to 2840 Hz/px. Since the slice plane is encoded in Cartesian mode, the native acquisition was performed in the coronal plane with field-of-view outside the anterior and posterior chest edges to prevent aliasing. It should be noted that resolutions, voxel spacings, and fields of view are not identical in CT and MR images. In addition, modalities may have been taken at different points in the respiratory cycle. \\ 
All volumes are resampled to have a voxel size of 0.6 $\times$ 0.6 $\times$ 0.6 mm$^3$. To align CT images with MR images, we first perform a rigid translation, followed by a deformable registration process using the EVolution algorithm proposed by \cite{EVolution2016}. This method incorporates a similarity term that prioritizes the alignment of edges and employs a diffeomorphic transformation to preserve the CT volume topology. Subsequently, we normalize MR images to have zero mean and unit variance, and for CT images, we use intensity values from the foreground classes to compute the mean and standard deviation, clip values to the 0.5 and 99.5 percentiles, and then normalize by subtracting the mean and dividing by the standard deviation. For data splitting, we use 82 patients for training, 20 patients for validation, and 20 patients for testing.

In the lung MR to CT synthesis task, we established the baseline using the traditional L1 loss and a perceptual loss leveraging MedicalNet's embeddings. We evaluated these losses against our proposed AFP loss, using various pre-trained segmentation networks: TotalSegmentator (TotalSeg), NaviAirway (Navi), and Holistic Airway Lesions (HAL). These networks were chosen to reflect a broad range of anatomical targets, covering both coarse and fine structures. TotalSegmentator, a publicly available general-purpose model, provides segmentations of large-scale anatomical regions such as bones, organs, and lung lobes; to improve its performance, we fine-tuned and fused several of its checkpoints into a higher-resolution version using the nnU-Net pipeline. NaviAirway, trained on healthy lungs, focuses on fine airway and bronchial tree structures, which are essential for evaluating structural fidelity in airway reconstructions. In contrast, HAL captures pathological patterns such as bronchiectasis and peribronchial thickening, making it particularly relevant for diseased lungs. Additionally, we explored a hybrid embedding strategy combining NaviAirway and HAL (denoted as Navi + HAL), designed to leverage complementary information from both healthy and pathological cases within the airway region. This hybrid approach allowed us to evaluate whether the fusion of distinct anatomical priors could further enhance structural preservation in the synthesized CT volumes.

\subsubsection*{Pelvis MR to CT Synthesis}
For the synthesis of pelvis images, we used the open-access dataset from Task 1 of the SynthRAD2023 challenge from \cite{synthrad}. This dataset includes paired MR and CT images from 180 patients used for training, with an additional 30 patients designated for validation. The dataset provides high-quality, well-aligned MR and CT image pairs, which eliminates the need for additional registration processes. The voxel sizes in this dataset are already standardized to 2.5 × 1 × 1 mm$^3$, ensuring consistent spatial resolution across all images. Similar preprocessing steps from the lung dataset are applied, with z-score normalization for MR images and clipping CT values to the 0.5 and 99.5 percentiles, followed by normalization using the mean and standard deviation. For data splitting, we use 150 patients for training, 30 patients for validation, and 30 patients for testing.

Similarly to the lung task, in the pelvic image synthesis task, we established a baseline using the traditional L1 loss and perceptual loss using the MedicalNet model. Our AFP loss, leveraging embeddings from the TotalSegmentator (denoted TotalSeg) network, was evaluated against this baseline. Furthermore, we explored a hybrid approach that combines L1 loss with AFP loss (denoted L1 + AFP) to assess the benefits of retaining the global reconstruction L1 loss.

\subsection{Evaluation Metrics}
To comprehensively assess our synthesis models, we employ a dual approach combining intensity-based and task-specific metrics.

\subsubsection*{Intensity-based Metrics}
Traditional intensity-based metrics, such as Mean Absolute Error (MAE) and Structural Similarity Index (SSIM), provide a measure of global image synthesis quality but may not adequately capture the nuances required for specific medical tasks. Accurate delineation of boundaries between different tissues, identification of lesions, and precise measurement of anatomical structures for surgical planning or disease monitoring necessitate evaluation metrics based on the reconstruction of anatomical structures. Although qualitative assessments by expert radiologists are valuable, they are also very tedious and time-consuming.

\subsubsection*{Task-specific Metrics}
To capture the nuances required for specific medical tasks, we introduce a downstream task - the segmentation of anatomical structures - as a key evaluation method. This approach provides a more precise measure of the model's performance at the anatomical level and helps quantify the false positive rate, crucial for clinical applications. We use segmentation-based metrics like the Dice score (DSC) and the Normalized Surface Distance (NSD) \cite{nsd} to evaluate local reconstructions. Although DSC and NSD capture related information, they provide complementary insights:  DSC serves as the reference metric for segmentation overlap, while NSD includes a manually defined margin of error, enhancing its utility in analyzing fine structures and providing robustness against registration inaccuracies \cite{metrics}. In our study, the margin of error for NSD calculations is set at twice the voxel size: 1.2 mm for lungs and 2 mm for pelvis. To mitigate potential biases introduced by relying on pre-trained segmentation networks, we evaluate our synthesis models across multiple segmentation tasks and models. This ensures a comprehensive and balanced assessment of their performance. For both tasks, the segmentation of the real CT is defined as a silver-standard, given that the actual segmentation masks are not available. 
For each test case, we perform segmentation independently using each of the selected pre-trained models. Therefore, in the lung MR-to-CT synthesis setting, three different segmentation masks are generated per patient, one from each model. Evaluation metrics (DSC, NSD) are computed for each model's output separately, and we report the average score accross the patients from the test set.
For additional insight, we also provide the 95th percentile Hausdorff distance (HD95) in supplemental material A.

\paragraph{Lung MR to CT Synthesis}
For the lung region, we use segmentations from three pre-trained models:
\begin{enumerate}
    \item The public NaviAirway network \cite{navi}, focused on segmenting airways and bronchi.
    \item The private Holistic Airway Lesions (HAL, \cite{HAL}) model measures anatomical structures within the lungs, such as bronchiectasis and peribronchial thickening.
    \item The publicly available TotalSegmentator \cite{totalseg} models, which have been fine-tuned and fused into one comprehensive model. This network focuses on coarser segmentations such as lung lobes, bones, muscles, and organs outside the lungs.
\end{enumerate}

\paragraph{Pelvis MR to CT Synthesis}
For the pelvis region, we use segmentations from the TotalSegmentator model. This network provides segmentation of various tissues in the pelvic region, including bones (e.g., sacrum, hips), organs (e.g., bladder, prostate), and muscles (e.g., iliopsoas, gluteus). 

This dual approach of intensity-based and task-specific metrics allows for a comprehensive evaluation of both global image quality and specific anatomical structure preservation in our synthesis models.

\section{Results}
\subsection{Lung MR to CT Synthesis}
\begin{figure*}[!ht]
\centering
\begin{tabular}{@{\hspace{0mm}}cccc@{\hspace{0mm}}}
\includegraphics[width=0.2\textwidth]{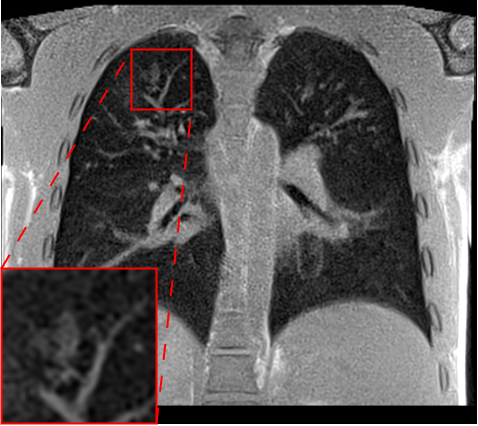}&
\includegraphics[width=0.2\textwidth]{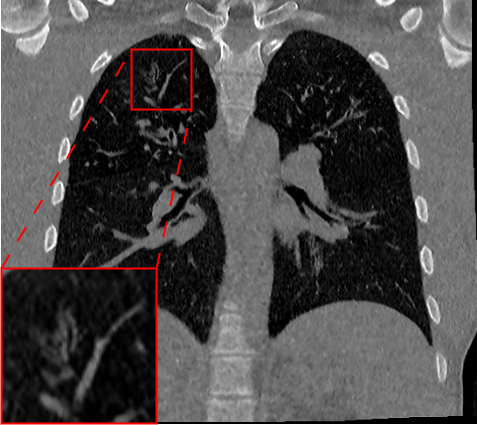}&
\includegraphics[width=0.2\textwidth]{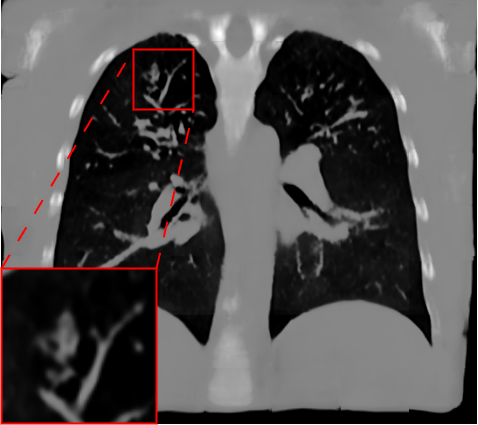}&
\includegraphics[width=0.2\textwidth]{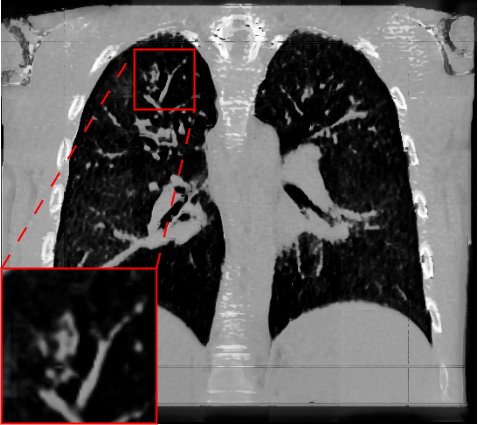} \\
Input MR & Real CT & L1 loss & Perceptual loss \\
\includegraphics[width=0.2\textwidth]{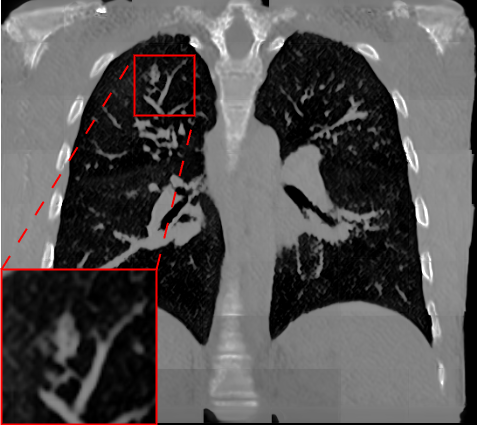}&
\includegraphics[width=0.2\textwidth]{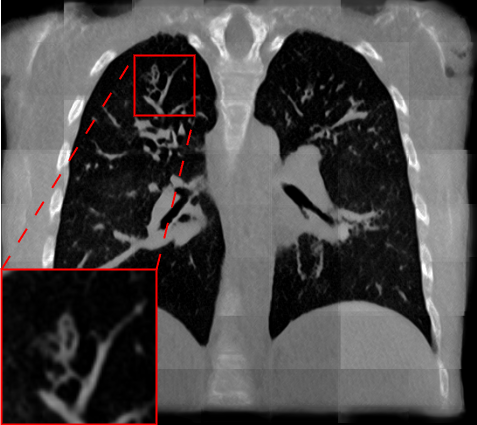}&
\includegraphics[width=0.2\textwidth]{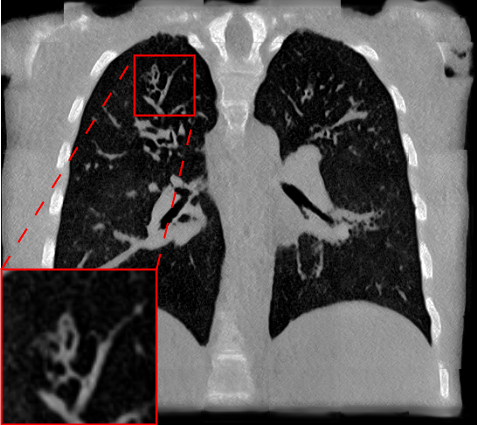}&
\includegraphics[width=0.2\textwidth]{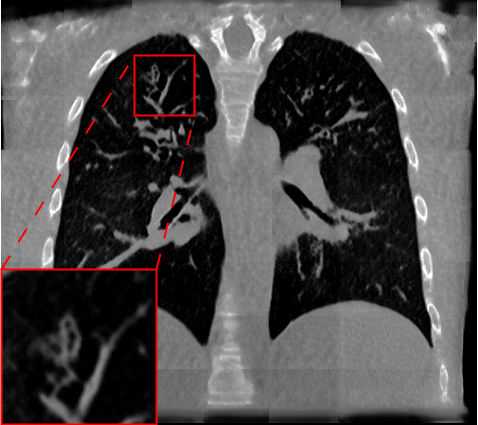} \\
AFP loss w/ & AFP loss w/ & AFP loss w/ & AFP loss w/ \\
TotalSegmentator & NaviAirway & HAL & NaviAirway + HAL
\end{tabular}
\caption{Comparison between input MR image, ground truth CT, and synthesized CT from adapted nnU-Net with several loss functions. Coronal slices are presented, including a zoomed-in view of the bronchi reconstruction. The zoomed-in view demonstrates that AFP loss with NaviAirway and HAL features enables the most faithful reconstruction of the real CT.}
\label{fig:slice_lungs}
\end{figure*}

Fig. \ref{fig:slice_lungs} presents a visual comparison of lung coronal sections from different models. The model using L1 loss results in blurred images, with particularly poor bone reconstruction. In contrast, the MedicalNet-based model produces sharper images but introduces artifacts in anatomical areas. Models based on AFP loss offer visually appealing results, with well-reconstructed anatomical bones and lung tissue. Additionally, only NaviAirway and HAL methods provide detailed bronchial reconstruction, as highlighted in the zoomed-in section. 

To handle memory constraints during training and inference, we used a patch-based strategy. We experimented with multiple patch sizes, including $64\times 64 \times 64$ and $128 \times 128 \times 128$, and also used nnU-Net's default adaptive windowing strategy (e.g. $128 \times 112 \times 160$ for thoracic cases). While larger patches provide more contextual information and generally improve reconstruction quality, the overall performance did not differ substantially from nnU-Net’s dynamic patching. Some visible artifacts in Fig.\ref{fig:slice_lungs} result from patch combination, especially in regions not covered by the segmentation network used for AFP supervision. For instance, when the AFP loss relies on the NaviAirway model, artifacts often appear outside the lung fields, as the network lacks anatomical guidance in those regions. However, these artifacts are purely visual and do not affect the downstream segmentation results or the quantitative metrics reported, which are detailed in the upcoming paragraph.

\begin{table*}[!h]
\centering
\caption{Comparison of 3D lung networks performances using MAE and SSIM between real CT and synthesized CT, as well as Dice score and NSD on airways segmentation using the NaviAirway pipeline. Results include adapted nnU-Net model with various losses and GAN-based SPADE model. Performance values within each metric column are color-coded for visual comparison: dark green (optimal), green (favorable), orange (moderate), and red (suboptimal).}
\begin{footnotesize}
\begin{tabular}{l|c|c||c|c}
 & \multicolumn{2}{c||}{\textbf{Intensity-based metrics}} & \multicolumn{2}{c}{\textbf{Task-based metrics}} \\ \hline
\textbf{Model / Loss} & \textbf{MAE} & \textbf{SSIM} & \textbf{Dice} & \textbf{NSD} \\ \hline
nnU-Net / L1 & \cellcolor{1} $\mathbf{48.72 \pm 12.84}$ & \cellcolor{1} $\mathbf{0.837 \pm 0.015}$ & \cellcolor{3}$0.534 \pm 0.065$ & \cellcolor{3}$0.644 \pm 0.089$ \\
nnU-Net / Perceptual & \cellcolor{4} $69.66 \pm 19.75$ & \cellcolor{3} $0.814 \pm 0.016$ & \cellcolor{4}$0.471 \pm 0.064$ & \cellcolor{4}$0.591 \pm 0.074$ \\
nnU-Net / AFP TotalSeg & \cellcolor{3} $59.40 \pm 14.91$ & \cellcolor{2}$0.828 \pm 0.014$ & \cellcolor{3}$0.530 \pm 0.068$ & \cellcolor{3}$0.667 \pm 0.076$ \\
nnU-Net / AFP Navi & \cellcolor{4} $64.98 \pm 16.88$ & \cellcolor{2} $0.824 \pm 0.014$ & \cellcolor{1}$0.580 \pm 0.059\textsuperscript{*}$ & \cellcolor{1}$0.720 \pm 0.064\textsuperscript{*}$ \\
nnU-Net / AFP HAL & \cellcolor{3}$60.92 \pm 15.90$ & \cellcolor{2}$0.827 \pm 0.014$ & \cellcolor{2}$0.564 \pm 0.067$ & \cellcolor{2}$0.702 \pm 0.079\textsuperscript{*}$ \\
nnU-Net / AFP Navi + HAL & \cellcolor{3} $61.47 \pm 15.98$ & \cellcolor{2}$0.827 \pm 0.014$ & \cellcolor{1}$\mathbf{0.584 \pm 0.063\textsuperscript{*}}$ & \cellcolor{1}$\mathbf{0.723 \pm 0.072\textsuperscript{*}}$ \\
\hline
SPADE / L1 & \cellcolor{2} $55.68 \pm 14.42$ & \cellcolor{2} $0.829 \pm 0.015$ & \cellcolor{3}$0.521 \pm 0.062$ & \cellcolor{3}$0.623 \pm 0.085$ \\
SPADE / AFP Navi & \cellcolor{4} $69.15 \pm 18.29$ & \cellcolor{3} $0.815 \pm 0.016$ & \cellcolor{2}$0.545 \pm 0.071$ & \cellcolor{2}$0.689 \pm 0.079\textsuperscript{*}$ \\ 
\end{tabular}
\end{footnotesize}
\footnotesize{Note: \textsuperscript{*}Indicates statistical significance compared to L1 loss, as measured using the Wilcoxon signed-rank test. P values less than 0.05 were considered statistically significant.}
\label{tab:mae_airways_lungs}
\end{table*}

\begin{figure*}[!h]
\centering
\begin{footnotesize}
\begin{tabular}{@{}c@{}}
\begin{tabular}{c@{\hspace{0.5cm}}c@{\hspace{0.5cm}}c@{\hspace{0.5cm}}c@{\hspace{0.5cm}}c}
\raisebox{-0.1cm}{\textcolor[rgb]{0.439, 0.596, 0.439}{\rule{0.4cm}{0.4cm}}} Airways &
\raisebox{-0.1cm}{\textcolor[rgb]{0.8, 0.9, 0.8}{\rule{0.4cm}{0.4cm}}} Trachea \\
\end{tabular}
\end{tabular}
\vspace{0.2cm} 

\begin{tabular}{@{\hspace{0mm}}cccccc@{\hspace{0mm}}}
\includegraphics[width=0.13\textwidth, trim=2cm 0cm 2cm 4cm, clip]{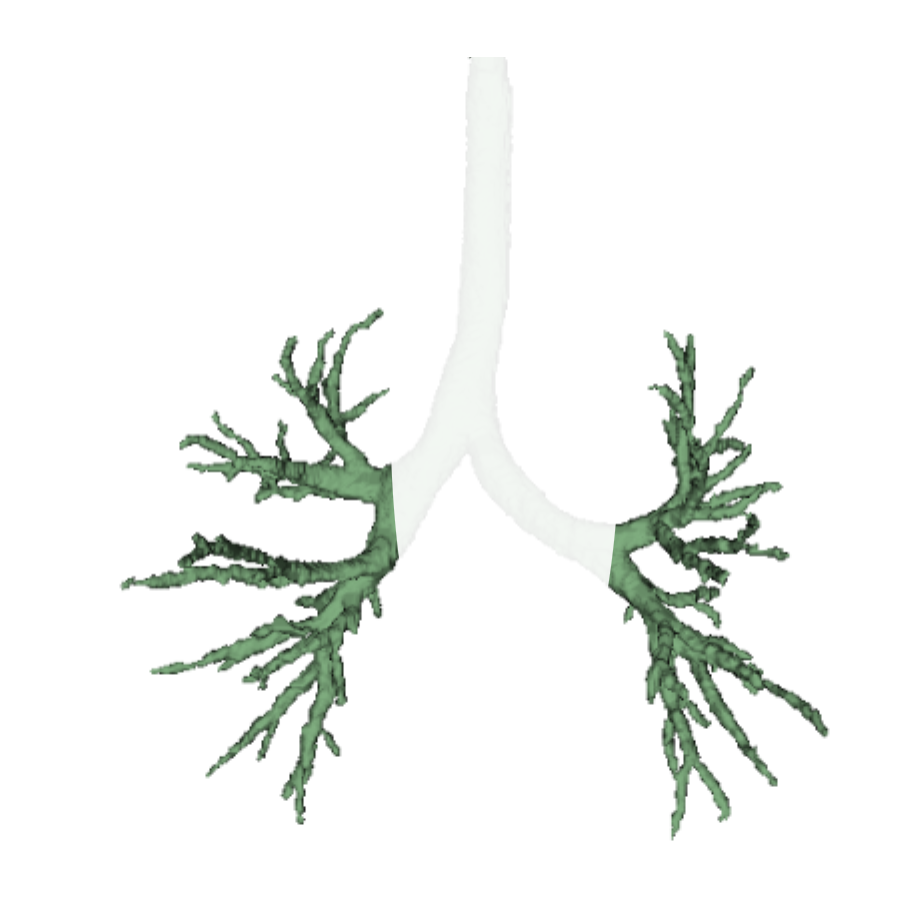}&
\includegraphics[width=0.13\textwidth, trim=2cm 0cm 2cm 4cm, clip]{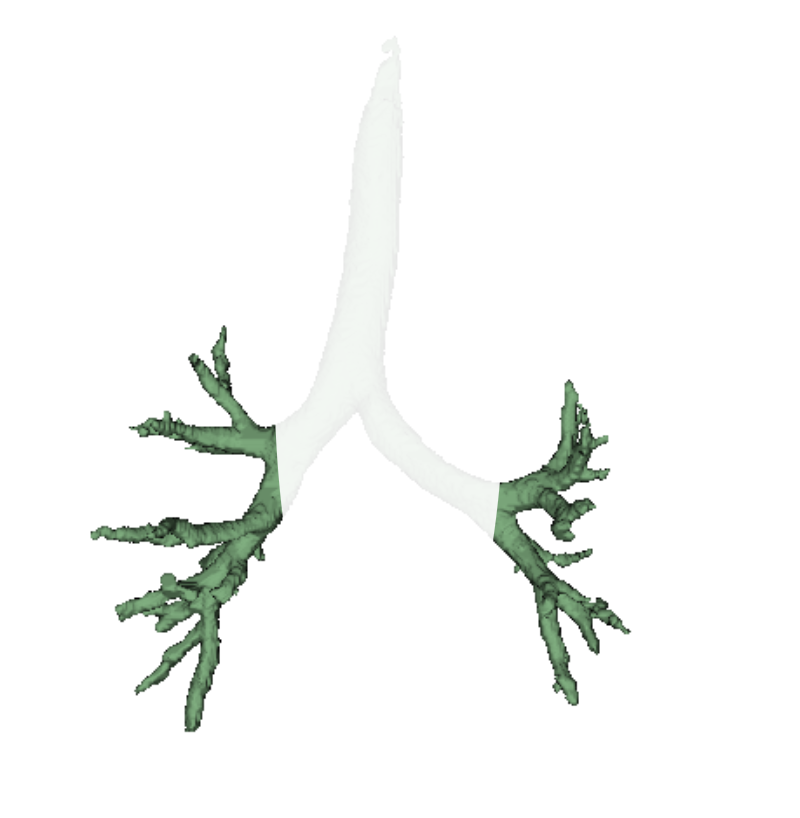}&
\includegraphics[width=0.13\textwidth, trim=2cm 0cm 2cm 4cm, clip]{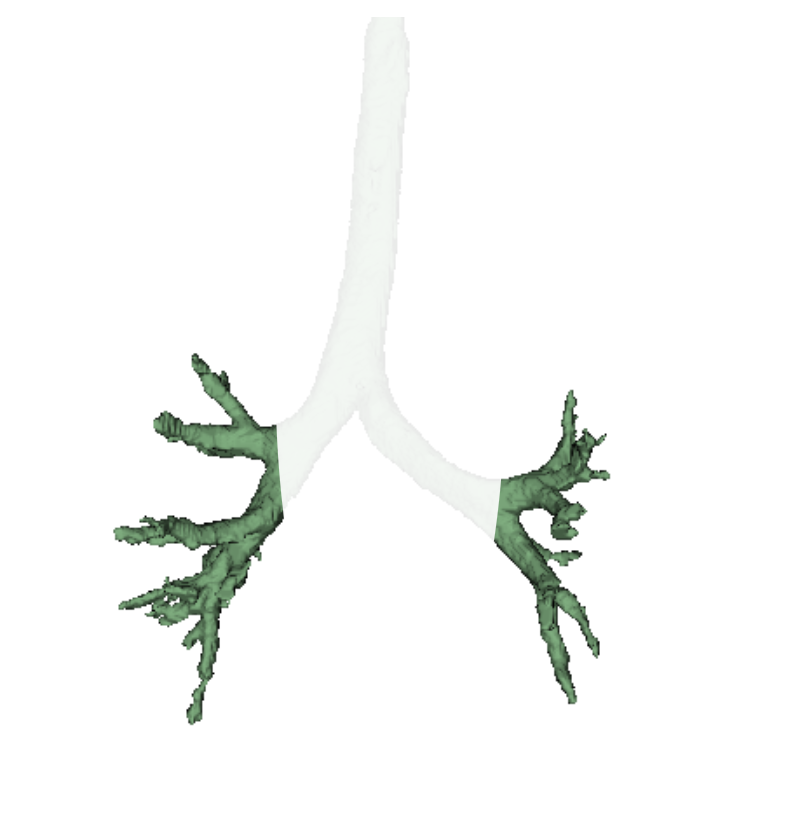}&
\includegraphics[width=0.13\textwidth, trim=2cm 0cm 2cm 4cm, clip]{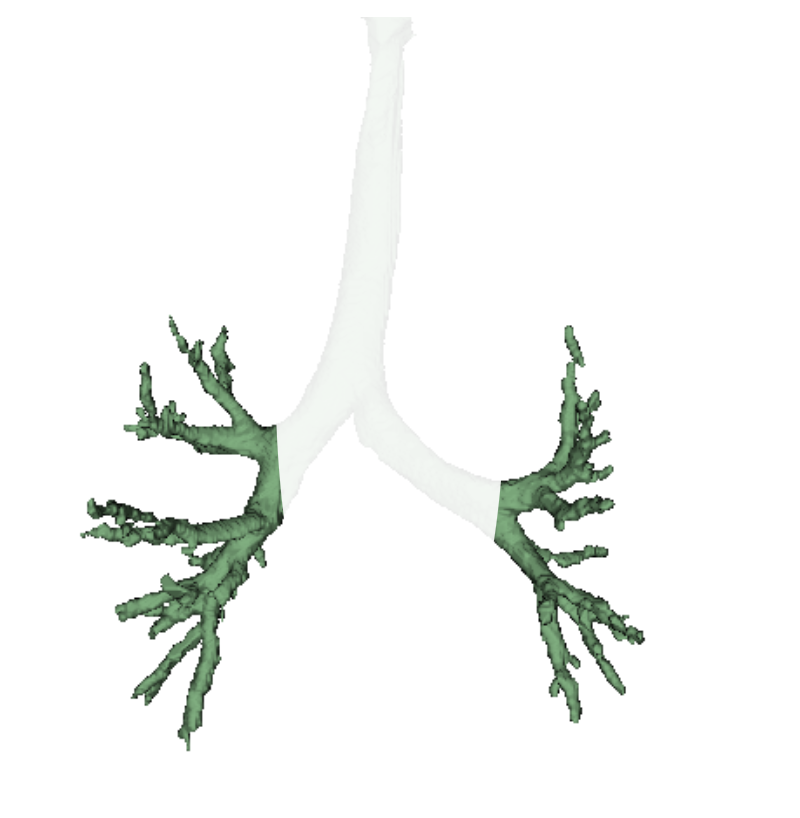}&
\includegraphics[width=0.13\textwidth, trim=2cm 0cm 2cm 4cm, clip]{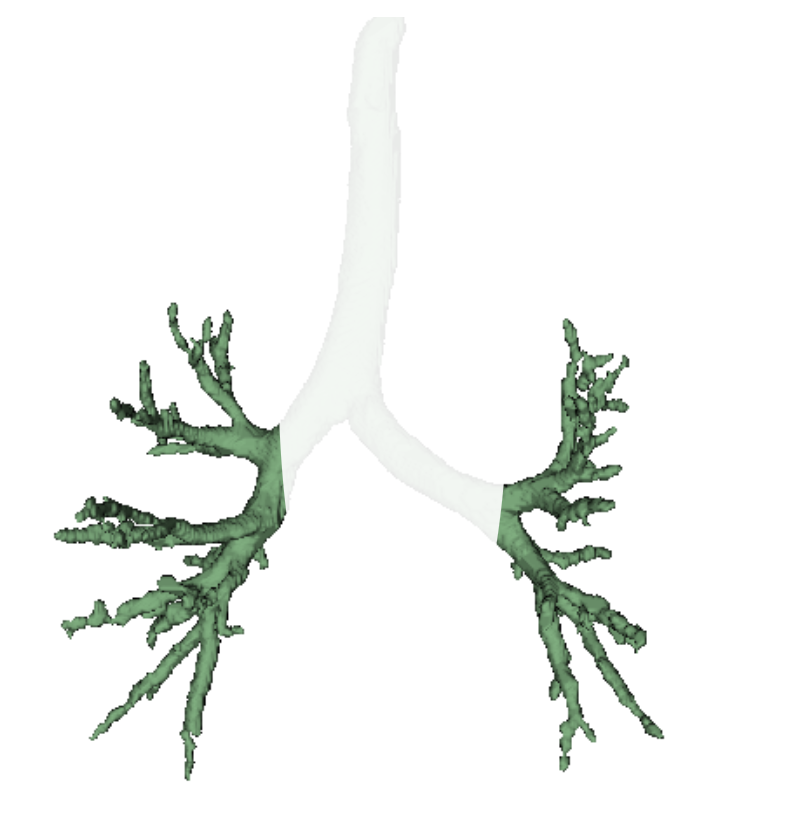}&
\includegraphics[width=0.13\textwidth, trim=2cm 0cm 2cm 4cm, clip]{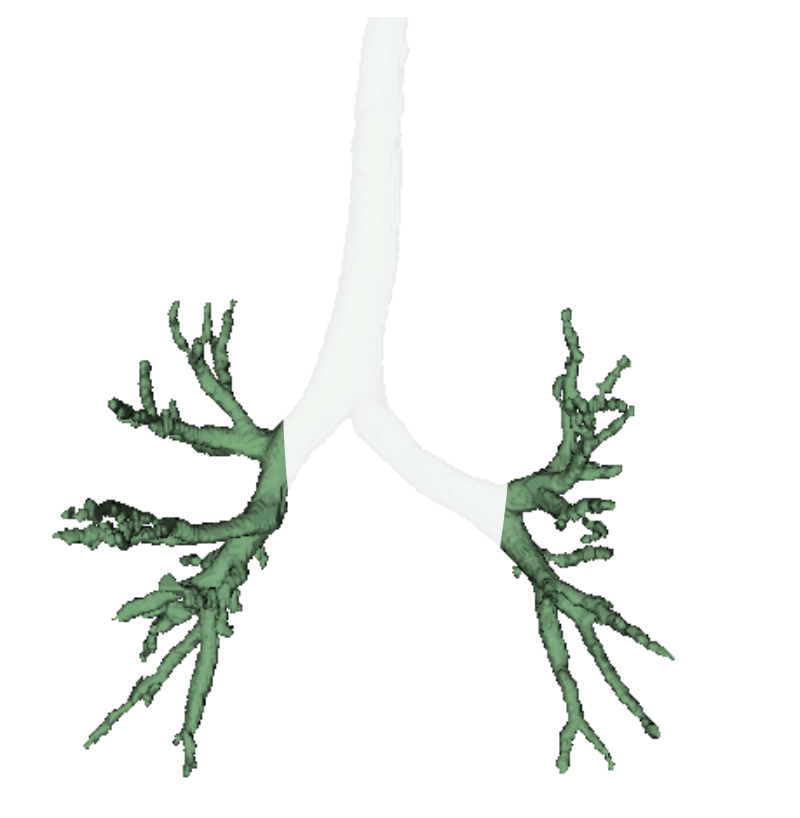} \\
Real CT & L1 & Perceptual & AFP TotalSeg & AFP Navi & AFP Navi + HAL \\
\end{tabular}

\end{footnotesize}
\caption{Example of 3D airway segmentation using the NaviAirway pipeline. The trachea is excluded for the calculation of metrics. The AFP loss facilitates the reconstruction of higher bronchi compared to the L1 loss, as depicted in Table \ref{tab:mae_airways_lungs}.}
\label{fig:naviairway}
\end{figure*}

Table \ref{tab:mae_airways_lungs} presents a quantitative evaluation of the model's performance on lung MR to CT synthesis, based on the MAE, SSIM, Dice score, and NSD between synthesized and ground truth CT images. The adapted nnU-Net trained with L1 loss delivers the best performance on intensity-based metrics, yielding a MAE of 48.72 and an SSIM of 0.837. In contrast, other models based on perceptual loss or AFP loss achieve average MAE results but maintain competitive SSIM values, for example, the AFP loss with TotalSegmentator embeddings achieves an SSIM of 0.828. In the context of airway segmentations using the NaviAirway pipeline, the adapted nnU-Net models with AFP loss from NaviAirway and HAL's embeddings deliver the best performance, achieving the highest Dice score of 0.584 and NSD value of 0.723. Conversely, models employing L1, perceptual, or AFP loss with TotalSegmentator's embeddings yield poorer results, lacking precise bronchial reconstruction. The GAN-based SPADE method generally underperforms compared to nnU-Net, but adding AFP loss to SPADE enhances its performance. These metrics align with qualitative analysis from Fig. \ref{fig:slice_lungs} and Fig. \ref{fig:naviairway}, with the models using AFP loss delivering the best performance in airway reconstruction.

\begin{figure*}[!h]
\centering
\begin{footnotesize}
\begin{tabular}{@{}c@{}}
\begin{tabular}{c@{\hspace{0.5cm}}c@{\hspace{0.5cm}}c@{\hspace{0.5cm}}c@{\hspace{0.5cm}}c@{\hspace{0.5cm}}c}
\raisebox{-0.1cm}{\textcolor[rgb]{0.549, 0.376, 0.314}{\rule{0.4cm}{0.4cm}}} Bronchial Mucus &
\raisebox{-0.1cm}{\textcolor[rgb]{0.933, 0.827, 0.561}{\rule{0.4cm}{0.4cm}}} Peribronchial Thickening &
\raisebox{-0.1cm}{\textcolor[rgb]{0.49, 0.812, 0.486}{\rule{0.4cm}{0.4cm}}} Bronchiectasis \\
\end{tabular}
\end{tabular}
\vspace{0.2cm} 

\begin{tabular}{@{\hspace{0mm}}cccccc@{\hspace{0mm}}}
\includegraphics[width=0.13\textwidth, trim=2cm 2cm 2cm 2cm, clip]{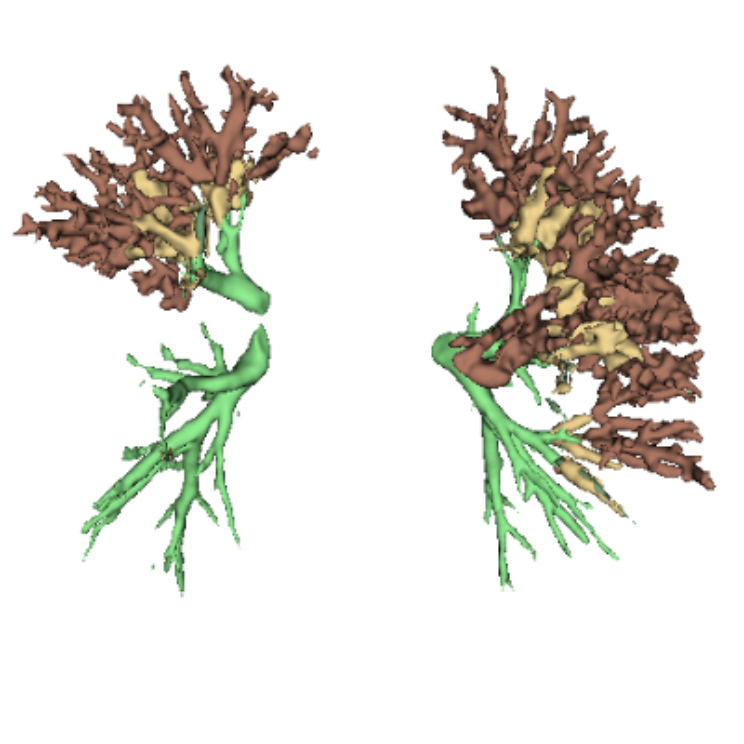}&
\includegraphics[width=0.13\textwidth, trim=2cm 2cm 2cm 2cm, clip]{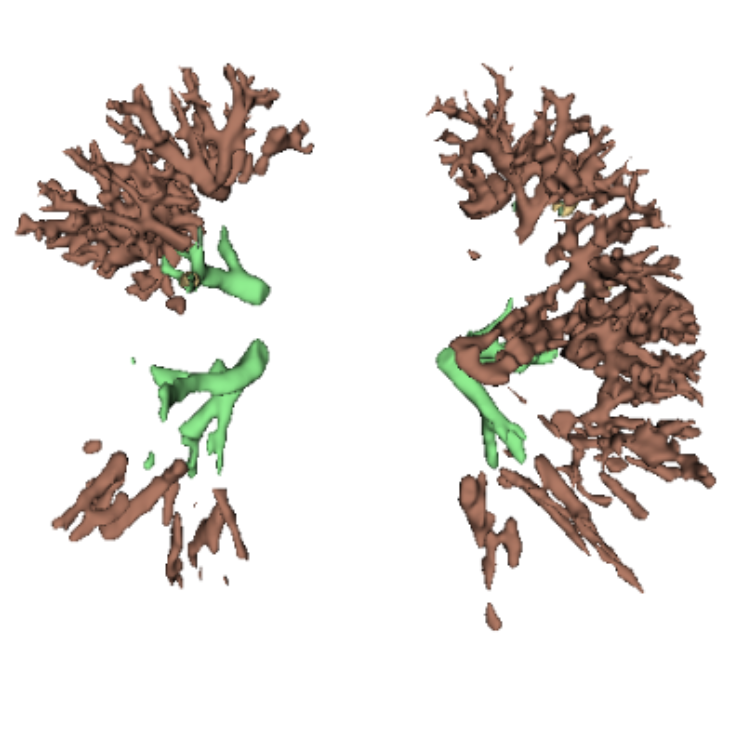}&
\includegraphics[width=0.13\textwidth, trim=2cm 2cm 2cm 2cm, clip]{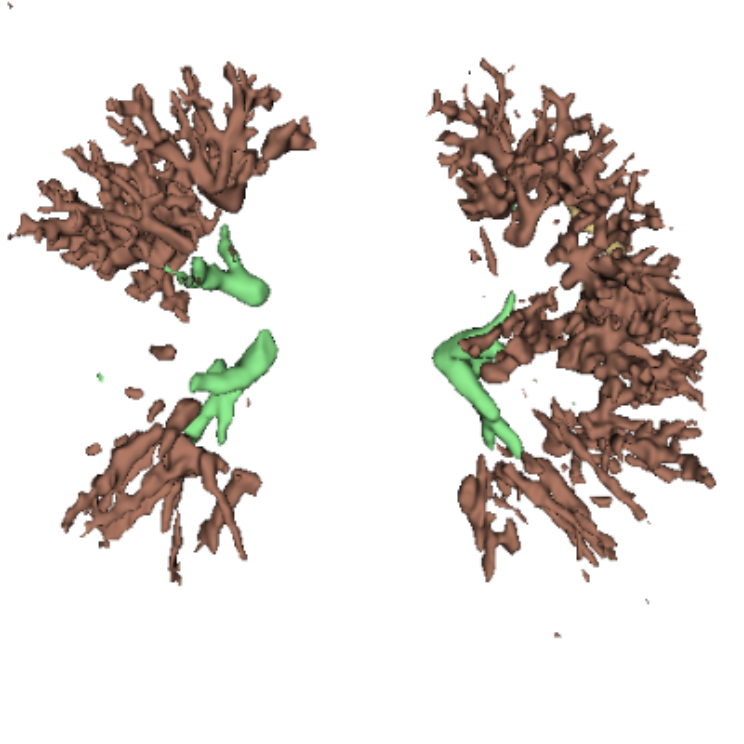}&
\includegraphics[width=0.13\textwidth, trim=2cm 2cm 2cm 2cm, clip]{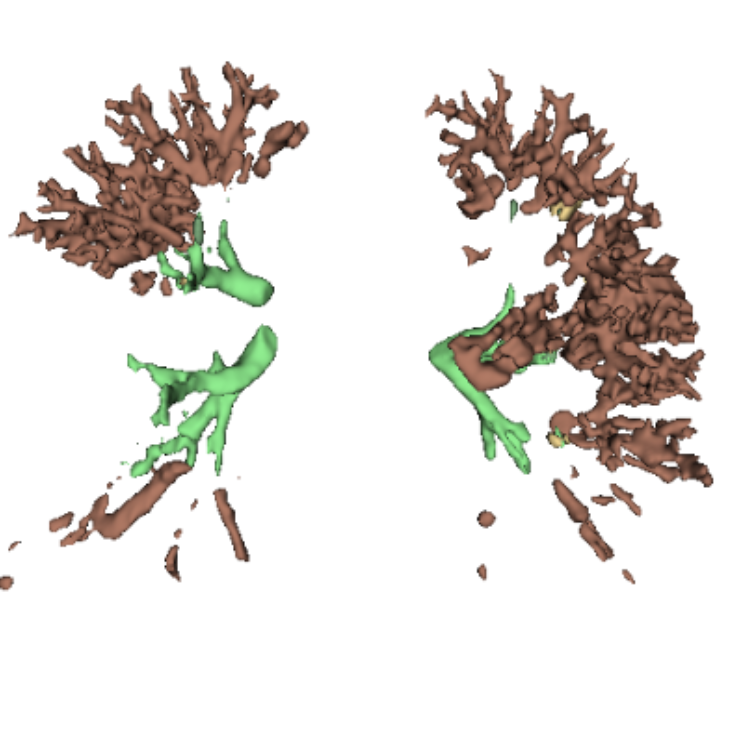}&
\includegraphics[width=0.13\textwidth, trim=2cm 2cm 2cm 2cm, clip]{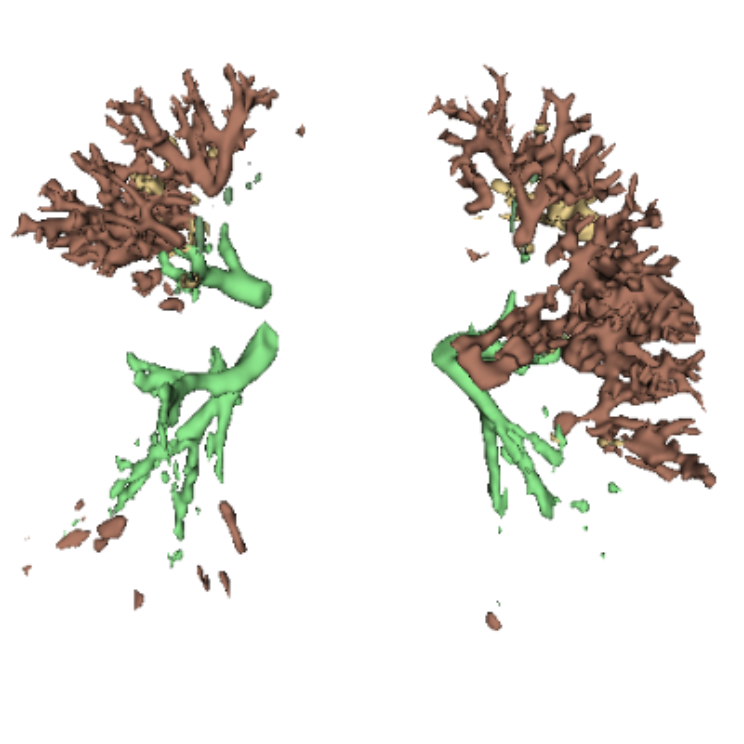}&
\includegraphics[width=0.13\textwidth, trim=2cm 2cm 2cm 2cm, clip]{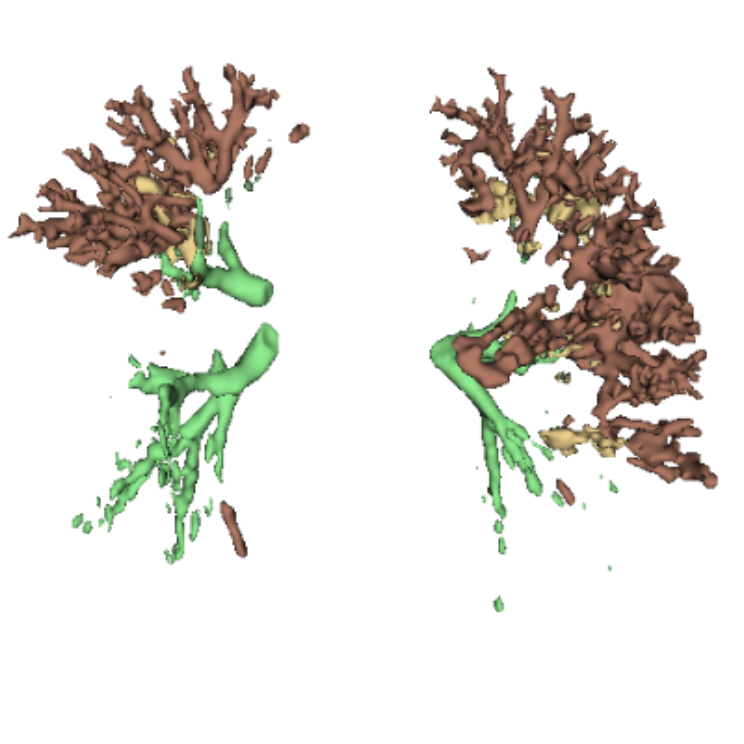} \\
Real CT & L1 & Perceptual & AFP TotalSeg & AFP Navi & AFP Navi + HAL \\
\end{tabular}

\end{footnotesize}
\caption{Comparison of airways lesions segmentations from real CT and synthesized CTs using the Holistic Airway Lesions (HAL) pipeline for an unhealthy patient. Bronchiectasis is represented in green, peribronchial thickenning in yellow and bronchial mucus in brown. Results correlate with Table \ref{tab:airway_lesions}, the AFP loss demonstrates the best visual outcomes, delivering fewer false positives in bronchial mucus compared to the L1 or perceptual losses.}
\label{fig:imene8}
\end{figure*}

Table \ref{tab:airway_lesions} and Fig. \ref{fig:imene8} provide a comparative evaluation of the models on  airway lesions segmentations using the Holistic Airway Lesions (HAL) pipeline with Dice score and NSD. Models leveraging the AFP loss using HAL's embeddings provide the best performances. Specifically, for bronchiectasis, the model using HAL's embeddings achieves an NSD of 0.516, significantly outperforming the L1 loss model which only reaches 0.373. Additionally, the model with AFP loss using only NaviAirway's embeddings also shows strong results, while other models are generally of poor quality on this task.

\begin{table*}[!h]
\centering
\caption{Comparison of 3D nnU-Net models with different losses for lung synthesis, evaluated using Dice score and NSD on Airway Lesions segmentations between real CT and synthesized CT.}
\begin{footnotesize}
\begin{tabular}{l|c|c|c|c}
\textbf{Loss} & \multicolumn{2}{c|}{\textbf{Bronchiectasis}} & \multicolumn{2}{c}{\textbf{Bronchial Mucus}} \\ \hline
 & \textbf{Dice} & \textbf{NSD} & \textbf{Dice} & \textbf{NSD} \\ \hline
L1 & \cellcolor{3}$0.287 \pm 0.087$ & \cellcolor{3}$0.373 \pm 0.108$ & \cellcolor{3}$0.281 \pm 0.094$ & \cellcolor{3}$0.369 \pm 0.146$ \\
Perceptual & \cellcolor{4}$0.237 \pm 0.087$ & \cellcolor{4}$0.308 \pm 0.110$ & \cellcolor{4}$0.225 \pm 0.095$ & \cellcolor{4}$0.276 \pm 0.132$ \\
AFP TotalSeg & \cellcolor{3}$0.308 \pm 0.086$ & \cellcolor{3}$0.376 \pm 0.119$ & \cellcolor{3}$0.303 \pm 0.094\textsuperscript{*}$ & \cellcolor{3}$0.375 \pm 0.146$ \\
AFP Navi & \cellcolor{2}$0.363 \pm 0.102\textsuperscript{*}$ & \cellcolor{2}$0.500 \pm 0.145\textsuperscript{*}$ & \cellcolor{2}$0.358 \pm 0.101\textsuperscript{*}$ & \cellcolor{2}$0.447 \pm 0.119\textsuperscript{*}$ \\
AFP HAL & \cellcolor{1}$\mathbf{0.393 \pm 0.105}\textsuperscript{*}$ & \cellcolor{1}$\mathbf{0.516 \pm 0.127}\textsuperscript{*}$ & \cellcolor{2}$\mathbf{0.363 \pm 0.109}\textsuperscript{*}$ & \cellcolor{1}$0.472 \pm 0.125\textsuperscript{*}$ \\
AFP Navi + HAL & \cellcolor{1}$0.388 \pm 0.096\textsuperscript{*}$ & \cellcolor{2}$0.496 \pm 0.122\textsuperscript{*}$ & \cellcolor{2}$0.362 \pm 0.099\textsuperscript{*}$ & \cellcolor{1}$\mathbf{0.484 \pm 0.134}\textsuperscript{*}$ \\
\end{tabular}
\end{footnotesize}

\footnotesize{Note: \textsuperscript{*}Indicates statistical significance compared to L1 loss, as measured using the Wilcoxon signed-rank test. P values less than 0.05 were considered statistically significant.}
\label{tab:airway_lesions}
\end{table*}

\begin{figure*}[!h]
\centering
\begin{footnotesize}
\begin{tabular}{@{}c@{}}
\begin{tabular}{c@{\hspace{0.5cm}}c@{\hspace{0.5cm}}c@{\hspace{0.5cm}}c@{\hspace{0.5cm}}c@{\hspace{0.5cm}}c}
\raisebox{-0.1cm}{\textcolor[rgb]{0.933, 0.827, 0.561}{\rule{0.4cm}{0.4cm}}} Bones &
\raisebox{-0.1cm}{\textcolor[rgb]{0.549, 0.376, 0.314}{\rule{0.4cm}{0.4cm}}} L. Upper &
\raisebox{-0.1cm}{\textcolor[rgb]{0.388, 0.651, 0.745}{\rule{0.4cm}{0.4cm}}} L. Lower &
\raisebox{-0.1cm}{\textcolor[rgb]{0.765, 0.353, 0.278}{\rule{0.4cm}{0.4cm}}} R. Upper &
\raisebox{-0.1cm}{\textcolor[rgb]{0.835, 0.49, 0.38}{\rule{0.4cm}{0.4cm}}} R. Middle &
\raisebox{-0.1cm}{\textcolor[rgb]{0.49, 0.812, 0.486}{\rule{0.4cm}{0.4cm}}} R. Lower \\
\end{tabular}
\end{tabular}

\vspace{0.1cm} 
\begin{tabular}{@{\hspace{0mm}}cccccc@{\hspace{0mm}}}
\includegraphics[width=0.13\textwidth, trim=2cm 0cm 2cm 0cm, clip]{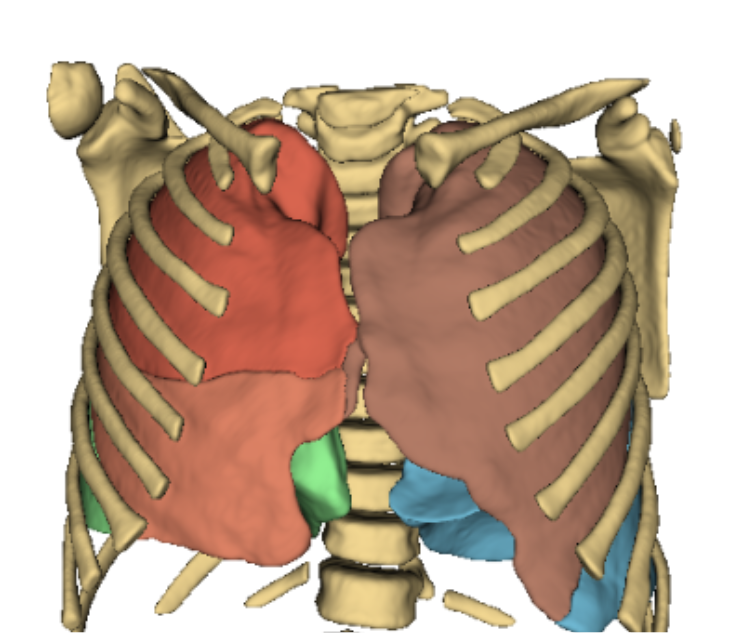} &
\includegraphics[width=0.13\textwidth, trim=2cm 0cm 2cm 0cm, clip]{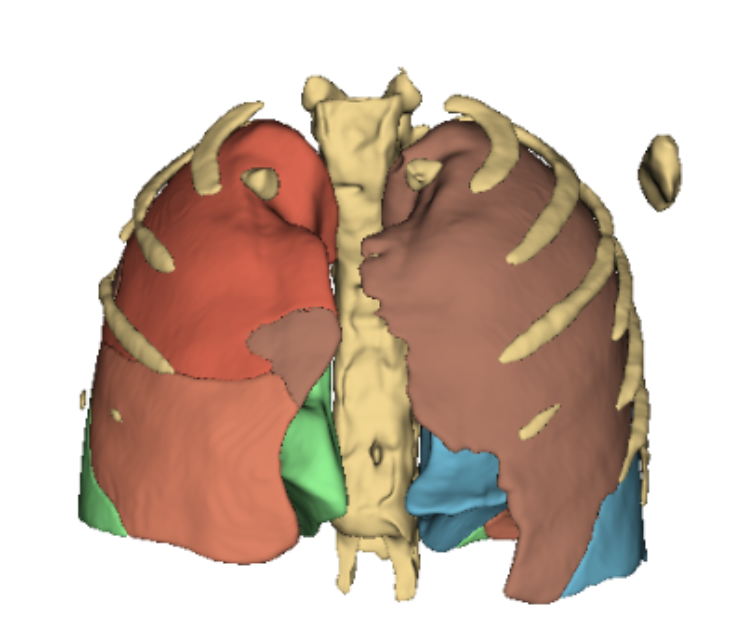} &
\includegraphics[width=0.13\textwidth, trim=2cm 0cm 2cm 0cm, clip]{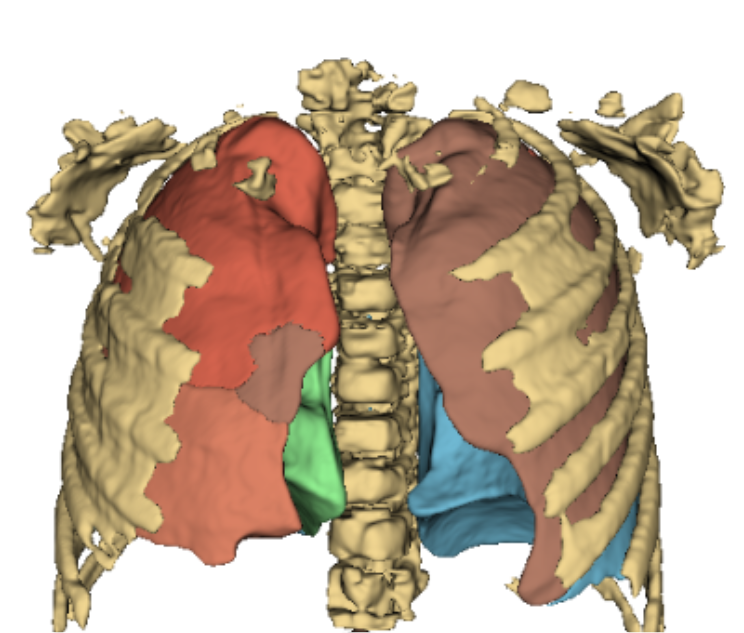} &
\includegraphics[width=0.13\textwidth, trim=2cm 0cm 2cm 0cm, clip]{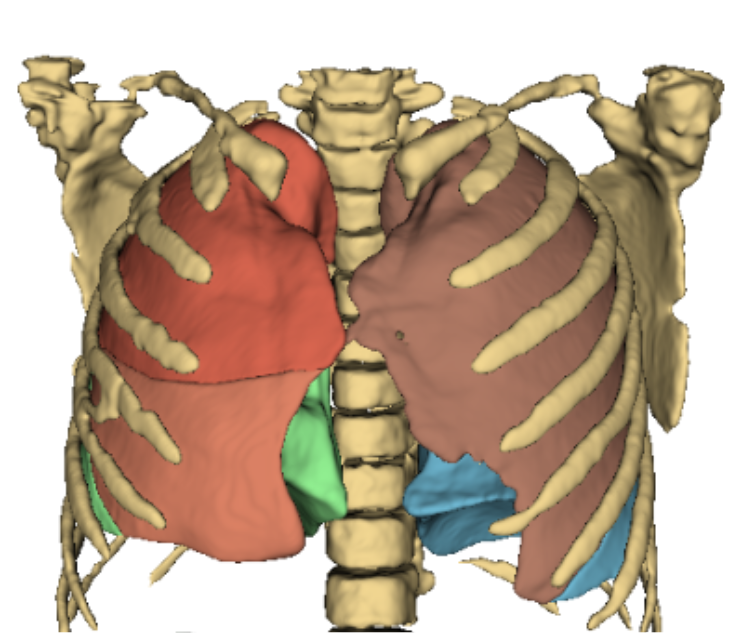} &
\includegraphics[width=0.13\textwidth, trim=2cm 0cm 2cm 0cm, clip]{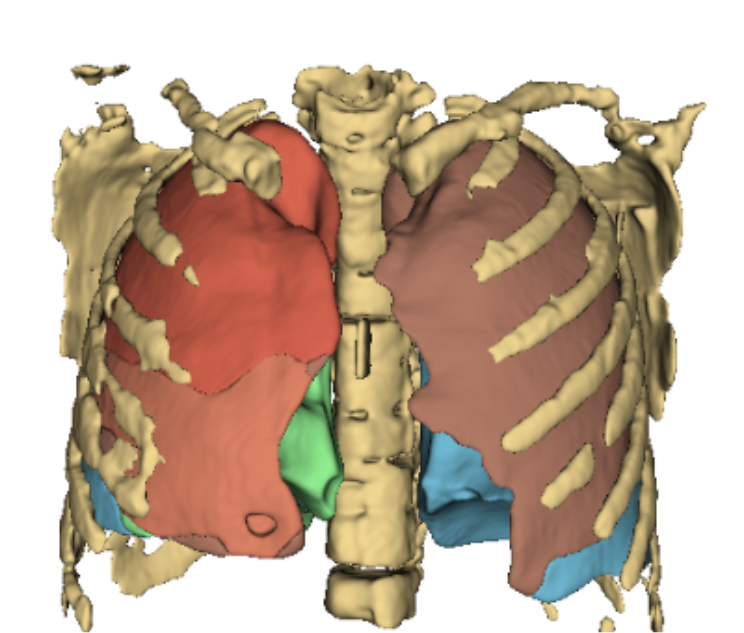} &
\includegraphics[width=0.13\textwidth, trim=2cm 0cm 2cm 0cm, clip]{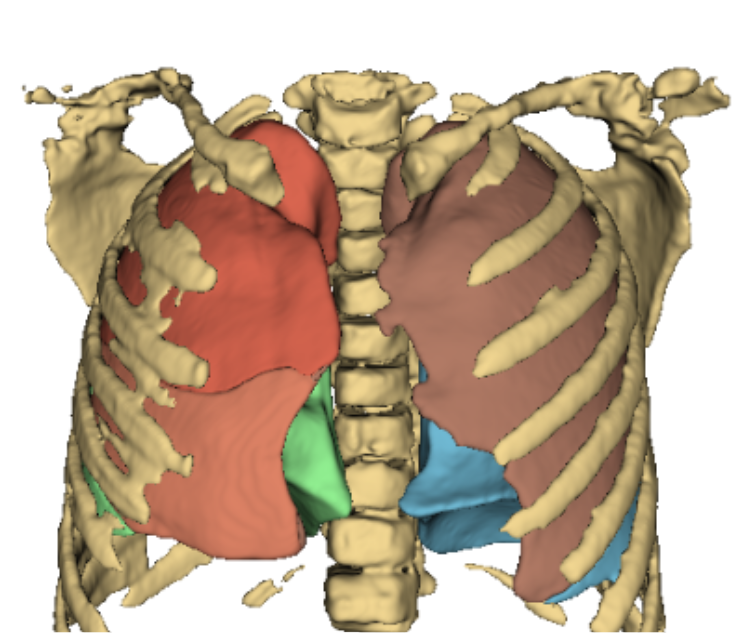} \\
Real CT & L1 & Perceptual & AFP TotalSeg & AFP NaviAirway & AFP Navi + HAL \\
\end{tabular}

\end{footnotesize}
\caption{Comparison of lung label segmentations from real CT and synthesized CTs using the TotalSegmentator pipeline (L: Left, R: Right, Upper/Middle/Lower Lobe).  Visual results align with Table~\ref{tab:totalseg_lungs}: the FP loss achieves superior reconstruction, particularly in the bones.}

\label{fig:totalseg_lungs}
\end{figure*}

Table \ref{tab:totalseg_lungs} and Fig. \ref{fig:totalseg_lungs} present a comparison of the models on the reconstruction of lung anatomical regions using the TotalSegmentator (TotalSeg) pipeline with Dice score and NSD. The model trained with the AFP loss using TotalSeg's embeddings provide the best performances. For example, when reconstructing bones, this model achieves a NSD of 0.629, outperforming the L1 loss model (NSD: 0.493) as well as perceptual loss (NSD: 0.499). A combination of AFP loss using NaviAirway and HAL's embeddings also delivers positive results, whereas other models exhibit notably poorer reconstructions especially in the bones. 

\begin{table*}[!h]
\centering
\caption{Comparison of 3D nnU-Net models with different losses for lung synthesis, evaluated using Dice score and NSD on TotalSegmentator segmentations between real CT and synthesized CT.}
\begin{footnotesize}
\begin{tabular}{l|c|c|c|c}
\textbf{Loss} & \multicolumn{2}{c|}{\textbf{Muscle/Organs}} & \multicolumn{2}{c}{\textbf{Bones}} \\ \hline
 & \textbf{Dice} & \textbf{NSD} & \textbf{Dice} & \textbf{NSD} \\ \hline
L1 & \cellcolor{2}$0.794 \pm 0.053$ & \cellcolor{2}$0.871 \pm 0.058$ & \cellcolor{4}$0.402 \pm 0.081$ & \cellcolor{4}$0.493 \pm 0.100$ \\
Perceptual & \cellcolor{3}$0.774 \pm 0.043$ & \cellcolor{2}$0.850 \pm 0.048$ & \cellcolor{4}$0.405 \pm 0.058$ & \cellcolor{4}$0.499 \pm 0.071$ \\
AFP TotalSeg & \cellcolor{1}$\mathbf{0.811 \pm 0.057}\textsuperscript{*}$ & \cellcolor{1}$\mathbf{0.900 \pm 0.063}\textsuperscript{*}$ & \cellcolor{1}$\mathbf{0.516 \pm 0.070}\textsuperscript{*}$ & \cellcolor{1}$\mathbf{0.629 \pm 0.086}\textsuperscript{*}$ \\
AFP Navi & \cellcolor{3}$0.751 \pm 0.063$ & \cellcolor{3}$0.801 \pm 0.081$ & \cellcolor{3}$0.479 \pm 0.074\textsuperscript{*}$ & \cellcolor{3}$0.583 \pm 0.092\textsuperscript{*}$ \\
AFP HAL & \cellcolor{2}$0.785 \pm 0.064$ & \cellcolor{2}$0.859 \pm 0.071$ & \cellcolor{3}$0.481 \pm 0.076\textsuperscript{*}$ & \cellcolor{3}$0.590 \pm 0.092\textsuperscript{*}$ \\
AFP Navi + HAL & \cellcolor{2}$0.786 \pm 0.060$ & \cellcolor{2}$0.844 \pm 0.066$ & \cellcolor{2}$0.499 \pm 0.074\textsuperscript{*}$ & \cellcolor{2}$0.604 \pm 0.090\textsuperscript{*}$ \\
\end{tabular}
\end{footnotesize}

\footnotesize{Note: \textsuperscript{*}Indicates statistical significance compared to L1 loss, as measured using the Wilcoxon signed-rank test. P values less than 0.05 were considered statistically significant.}
\label{tab:totalseg_lungs}
\end{table*}

To address the potential concerns regarding clinical deployment, we report the computational cost and training time associated with different loss functions in Table \ref{tab:computational_cost}. All experiments were conducted on a single NVIDIA RTX 2080 Ti GPU with 12 GB of VRAM, a patch size of $128 \times 112 \times 160$ and a batch size of 2. The baseline model with L1 loss was the least demanding, while using a perceptual loss with MedicalNet increased both memory usage and training time slightly. The AFP loss introduced the highest overhead due to feature extraction from a deep 5-stage nnU-Net segmentation model, resulting in longer training. Despite this, training remained feasible on widely available hardware. It is important to note that the training time and memory usage for AFP loss can be optimized by reducing the complexity of the segmentation network used for feature extraction, for example, using a 4-stage nnU-Net instead of the standard 5-stage model. Furthermore, inference time and deployment cost are not affected by the choice of loss function, as these auxiliary networks are only used during training. Therefore, all models share the same inference speed and resource requirements, which is crucial for real-time clinical applications.

\begin{table}[!h]
\centering
\caption{Computational cost and training time for 3D nnU-Net models using different losses, with a patch size of $128 \times 112 \times 160$, a batch size of 2, and a training of 1000 epochs. Experiments conducted on an NVIDIA RTX 2080 Ti (12 GB VRAM). Inference time is reported for a full volume of size (533, 373, 533).}
\begin{footnotesize}
\begin{tabular}{l|c|c||c|c}
& \multicolumn{2}{c||}{\textbf{Training Phase}} & \multicolumn{2}{c}{\textbf{Test Phase}} \\ \hline
\textbf{Model / Loss} & \textbf{VRAM} & \textbf{Training Time} & \textbf{VRAM} & \textbf{Inference Time} \\ \hline
nnU-Net / L1 & \cellcolor{1} 7390 MB & \cellcolor{1} 29h & \cellcolor{1} 2860 MB & \cellcolor{1} 30s \\
nnU-Net / Perceptual & \cellcolor{2} 9070 MB & \cellcolor{2} 33h & \cellcolor{1} 2860 MB & \cellcolor{1} 30s \\
nnU-Net / AFP TotalSeg & \cellcolor{3} 10750 MB & \cellcolor{3} 47h & \cellcolor{1} 2860 MB & \cellcolor{1} 30s \\
\end{tabular}
\end{footnotesize}
\label{tab:computational_cost}
\end{table}

\subsection{Pelvis MR to CT Synthesis}
Table \ref{tab:mae_pelvis} presents a quantitative evaluation of the model's performance on pelvic MR to CT synthesis, based on the MAE and SSIM between synthesized and ground truth CT images. Similarly to the lung results, the model trained with L1 loss delivers the best performance on intensity-based metrics. Other models based on perceptual loss or AFP loss alone achieve average results in terms of MAE but maintain reasonable SSIM values. Finally, the combination of L1 and AFP losses provides solid results on these intensity-based metrics, bringing it closer to the high standards set by L1 loss. 

\begin{table}[!h]
\centering
\caption{Comparison of 3D nnU-Net models with different losses for pelvic synthesis using MAE and SSIM between real CT and synthesized CT.}
\begin{footnotesize}
\begin{tabular}{l|c|c}
\textbf{Loss} & \textbf{MAE} & \textbf{SSIM} \\ \hline
\L1 & \cellcolor{1} $\mathbf{65.300 \pm 15.180}$ & \cellcolor{1} $\mathbf{0.852 \pm 0.035}$ \\
Perceptual & \cellcolor{4} $81.359 \pm 30.717$ & \cellcolor{3} $0.838 \pm 0.038$ \\
L1 + AFP TotalSeg & \cellcolor{2} $66.937 \pm 15.116$ & \cellcolor{2} $0.847 \pm 0.035$ \\
AFP TotalSeg & \cellcolor{4} $85.258 \pm 14.471$ & \cellcolor{3} $0.839 \pm 0.036$ \\ 
\end{tabular}
\end{footnotesize}
\label{tab:mae_pelvis}
\end{table}

\begin{figure*}[!h]
\centering
\begin{footnotesize}
\begin{tabular}{@{\hspace{0mm}}ccccccc@{\hspace{0mm}}}
\includegraphics[width=0.14\textwidth, trim= 2cm 0 2cm 0,clip]{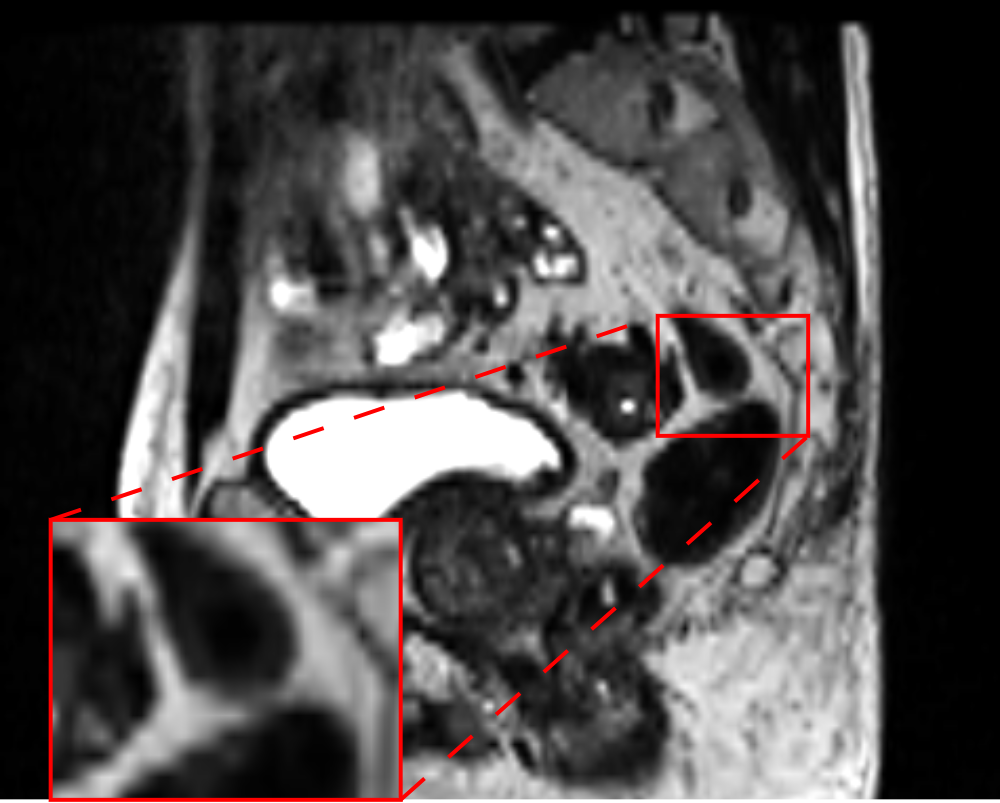}&
\includegraphics[width=0.14\textwidth, trim= 2cm 0 2cm 0,clip]{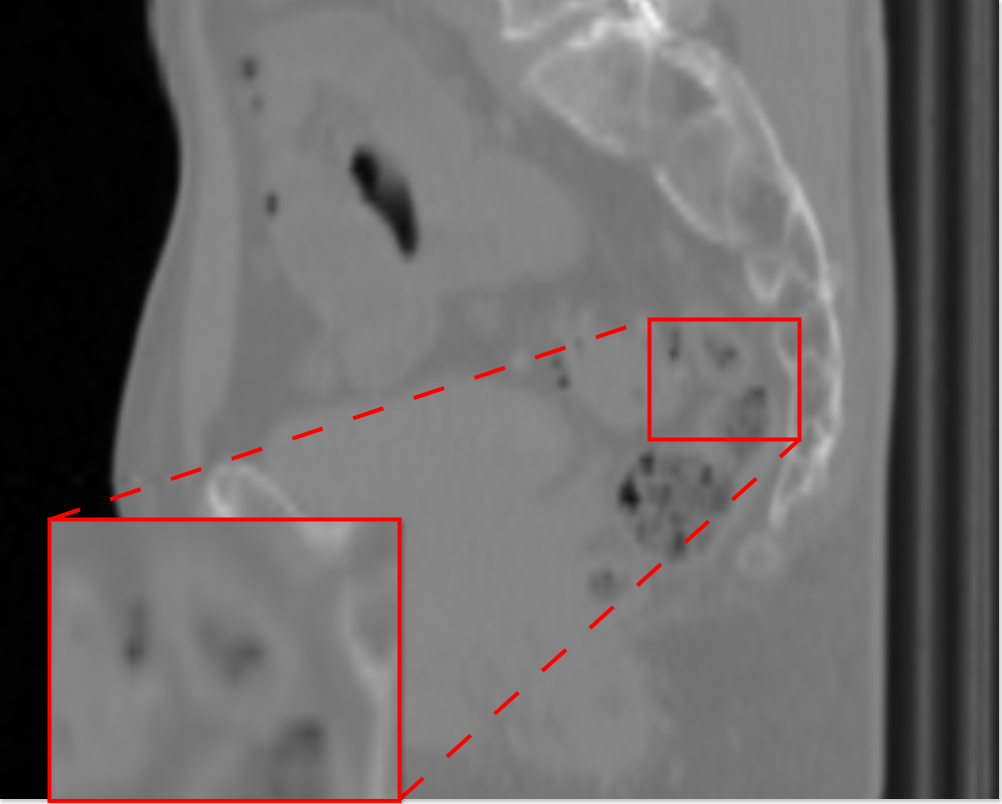}&
\includegraphics[width=0.14\textwidth, trim= 2cm 0 2cm 0,clip]{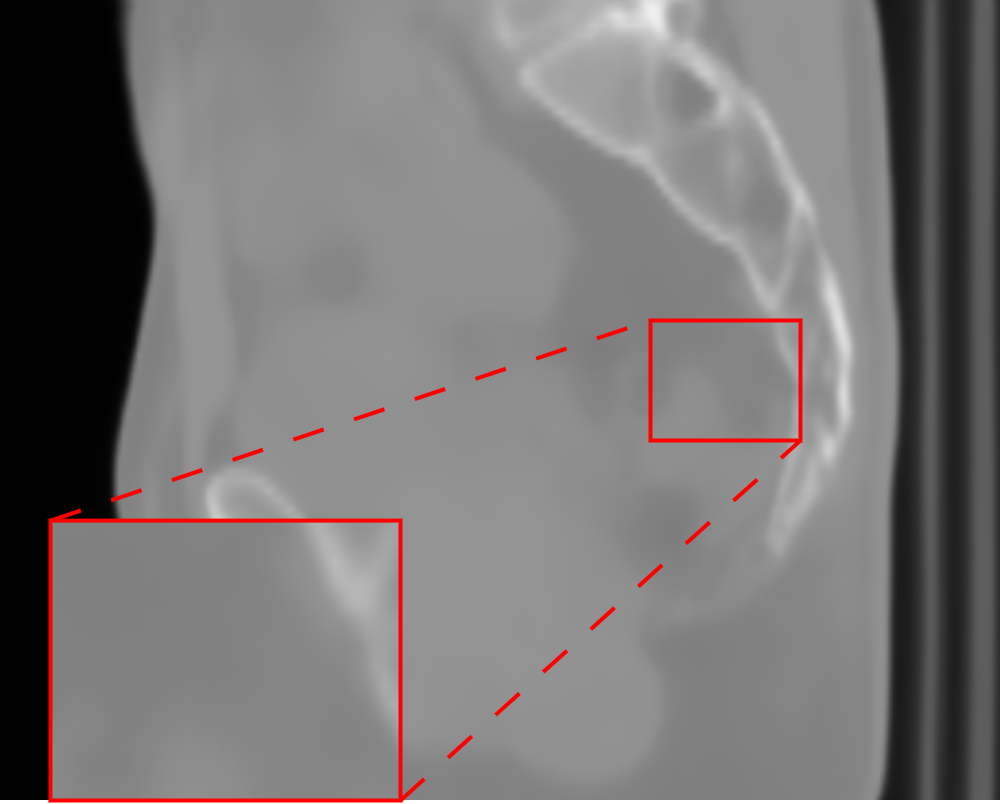}&
\includegraphics[width=0.14\textwidth, trim= 2cm 0 2cm 0,clip]{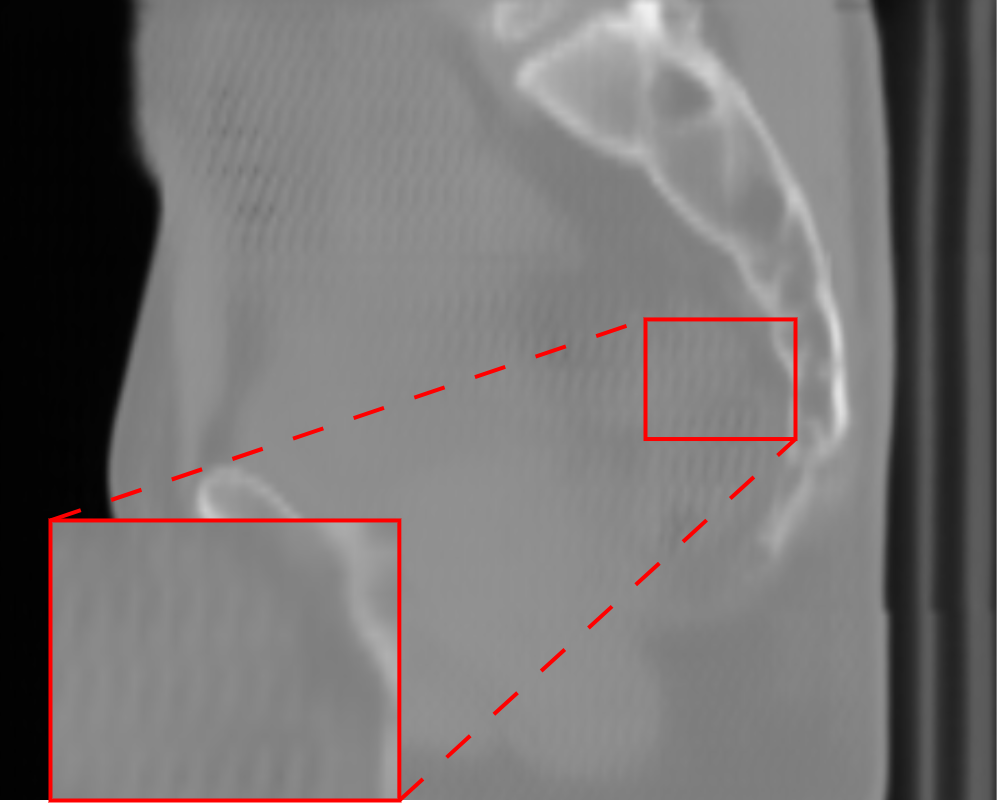}&
\includegraphics[width=0.14\textwidth, trim= 2cm 0 2cm 0,clip]{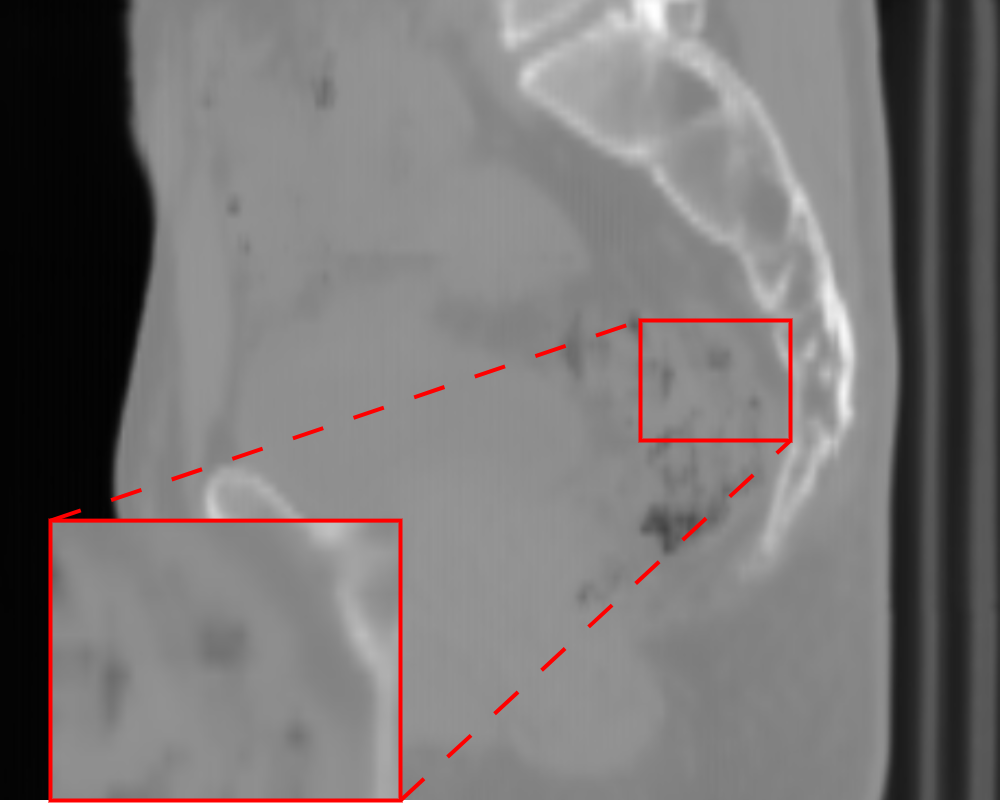} &
\includegraphics[width=0.14\textwidth, trim= 2cm 0 2cm 0,clip]{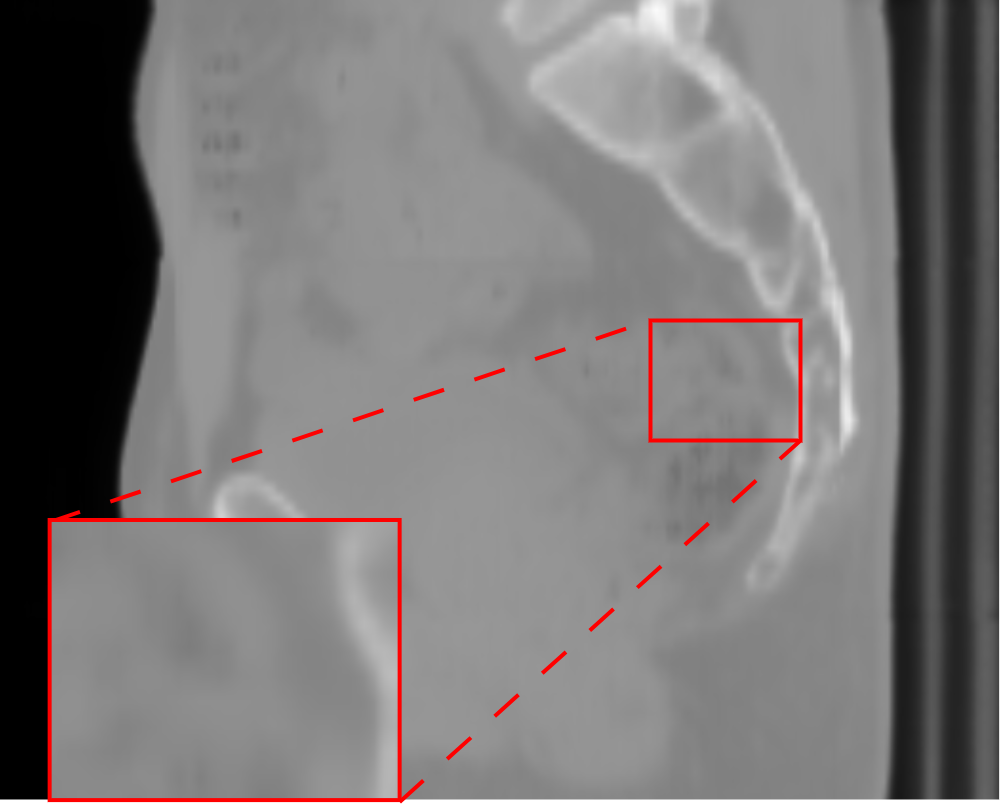} \\
Input MR & Real CT & L1 & Perceptual & AFP TotalSeg & L1 + AFP \\
& \includegraphics[width=0.14\textwidth, trim= 1cm 1cm 1cm 0cm,clip]{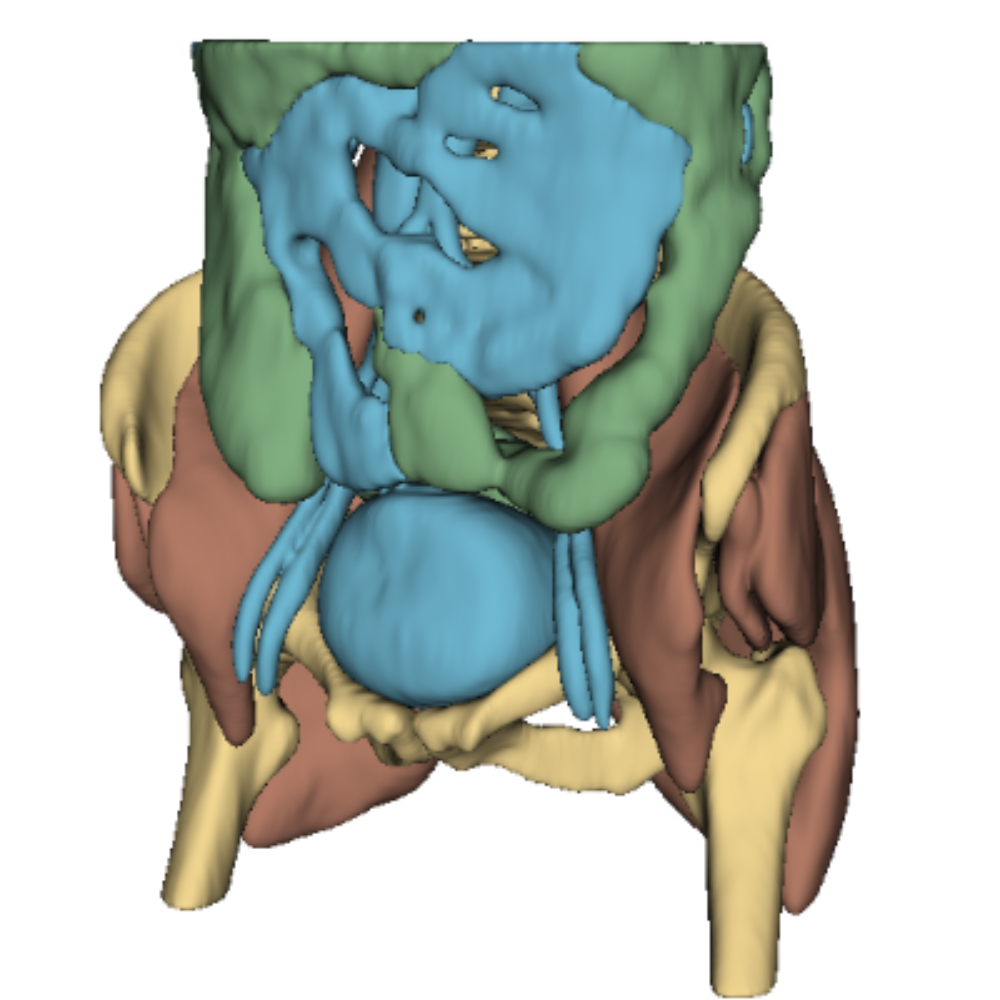}&
\includegraphics[width=0.14\textwidth, trim= 1cm 1cm 1cm 0cm, clip]{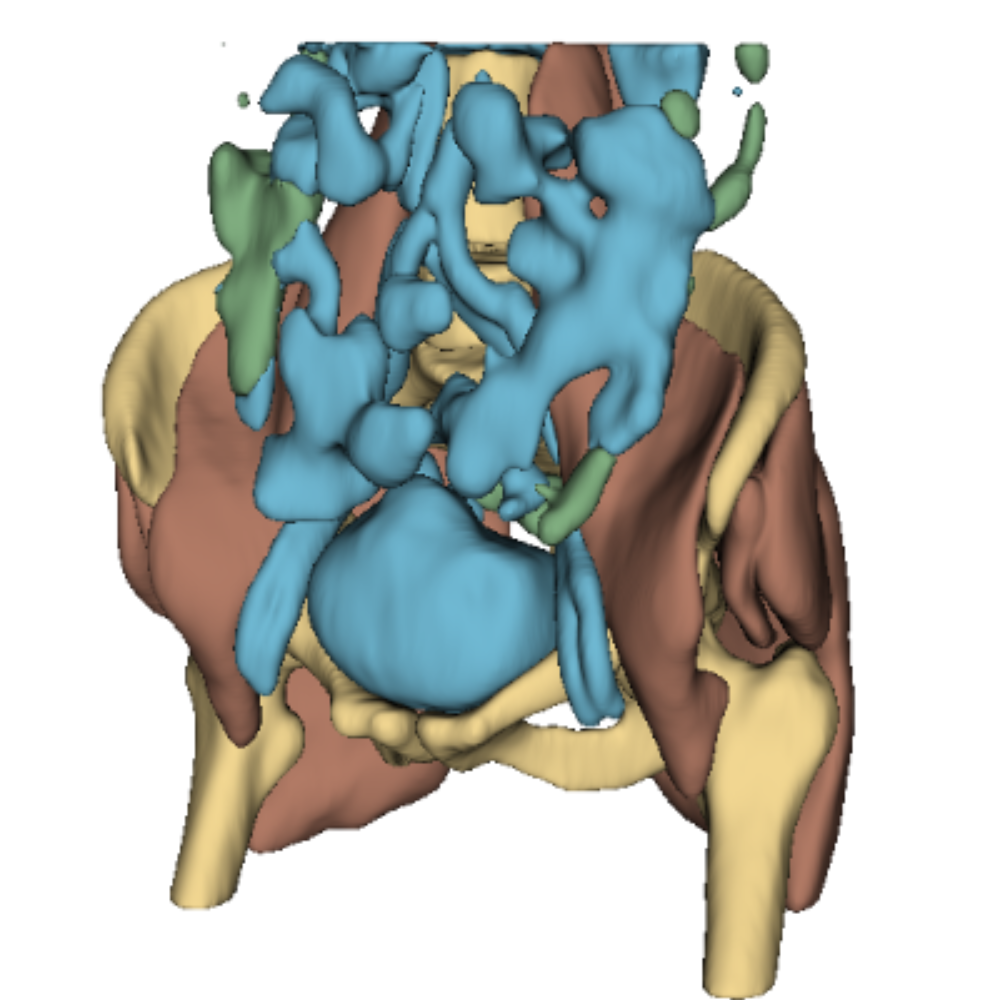}&
\includegraphics[width=0.14\textwidth, trim= 1cm 1cm 1cm 0cm,clip]{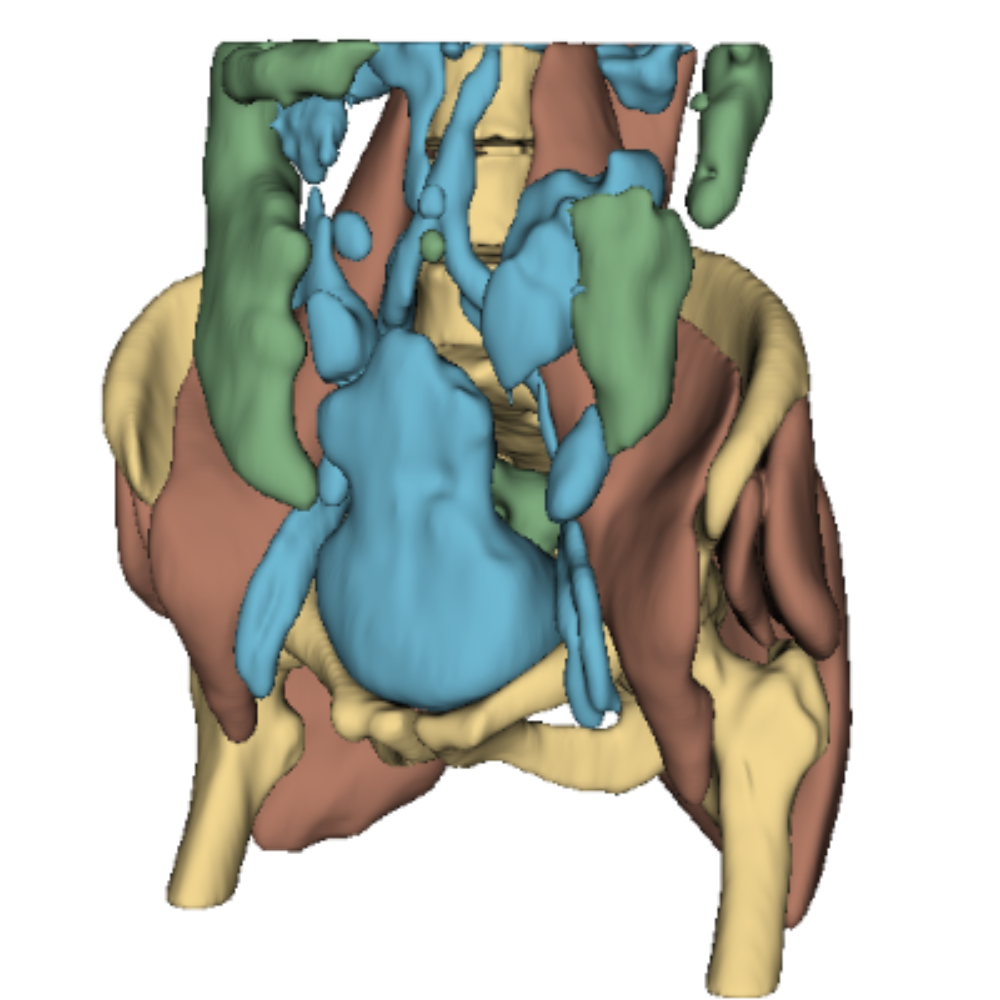}&
\includegraphics[width=0.14\textwidth, trim= 1cm 1cm 1cm 0cm,clip]{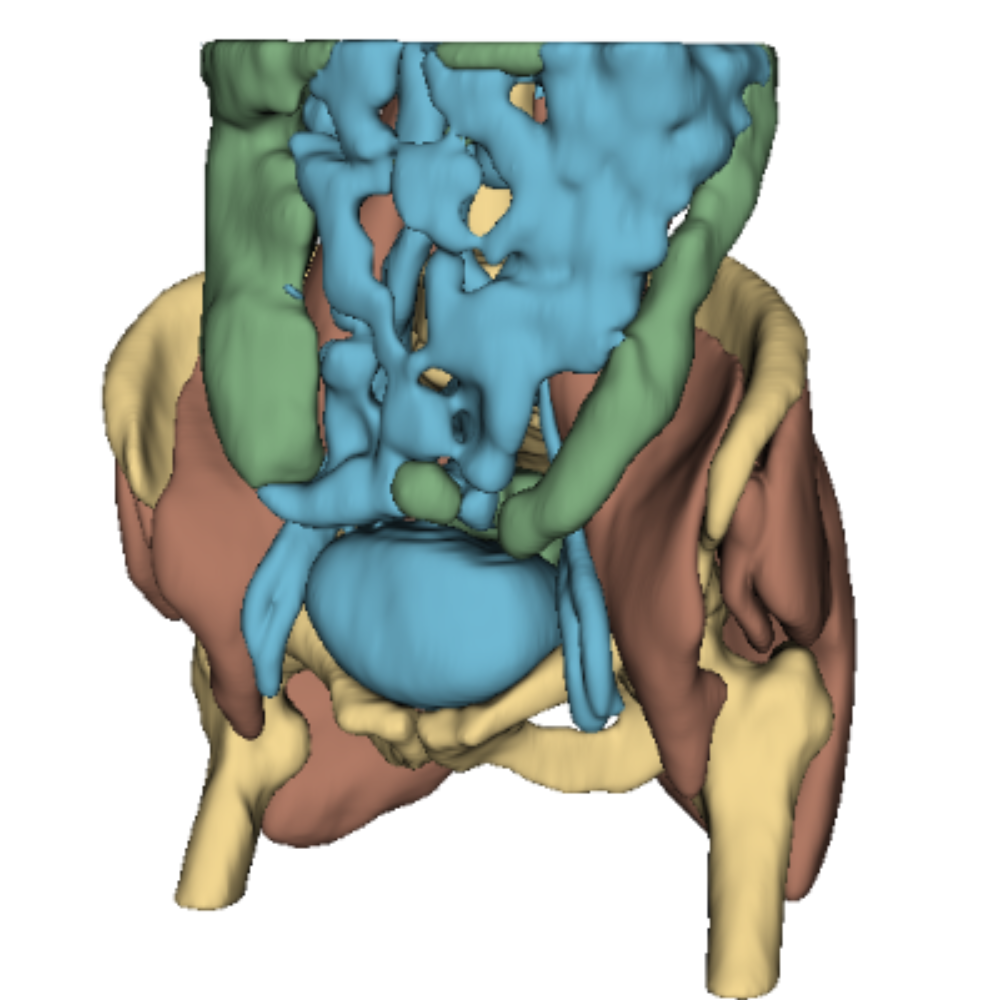} &
\includegraphics[width=0.14\textwidth, trim= 1cm 1cm 1cm 0cm,clip]{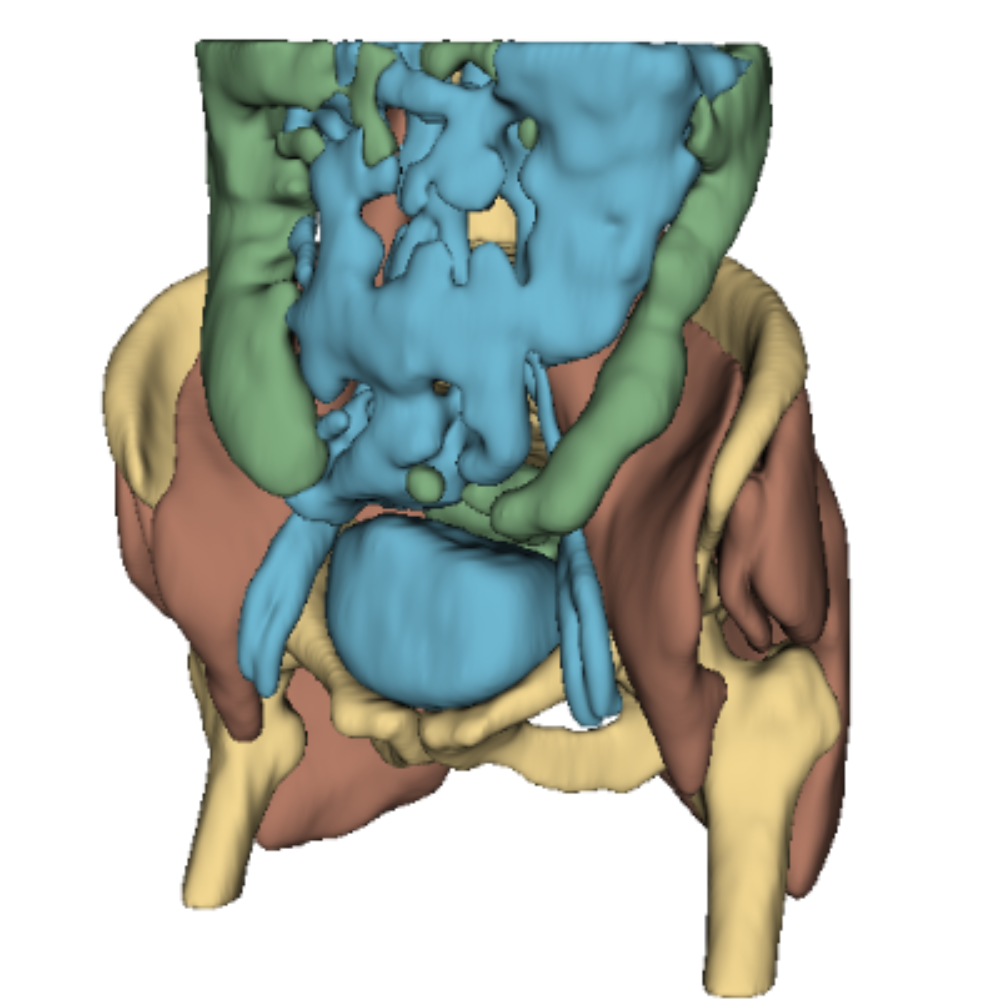} \\
\end{tabular}

\vspace{0.1cm}
\begin{tabular}{@{\hspace{2cm}}c@{\hspace{0.5cm}}c@{\hspace{0.5cm}}c@{\hspace{0.5cm}}c@{\hspace{0.5cm}}c@{\hspace{0.5cm}}c}
& \raisebox{-0.1cm}{\textcolor[rgb]{0.933, 0.827, 0.561}{\rule{0.4cm}{0.4cm}}} Bones &
\raisebox{-0.1cm}{\textcolor[rgb]{0.549, 0.376, 0.314}{\rule{0.4cm}{0.4cm}}} Muscles &
\raisebox{-0.1cm}{\textcolor[rgb]{0.388, 0.651, 0.745}{\rule{0.4cm}{0.4cm}}} Organs &
\raisebox{-0.1cm}{\textcolor[rgb]{0.493, 0.664, 0.492}{\rule{0.4cm}{0.4cm}}} Colon Region &
\end{tabular}

\end{footnotesize}
\caption{Comparison between input MR image, ground truth CT, and synthesized CT from nnU-Net translation with several loss functions. Sagittal slices are presented, including a zoomed-in view of the colon reconstruction on the top row, and 3D view of TotalSegmentator segmentations on the bottom row. Visual results align with Table \ref{tab:totalseg_pelvis} and show that AFP loss enables a better anatomical reconstruction, especially in the organs and colon. }
\label{fig:pelvis}
\end{figure*}

Fig. \ref{fig:pelvis} presents a visual comparison of pelvic sagittal sections from different models. The model trained with L1 loss displays blurry results, with smooth intensities across the body and a noticeable lack of precise structural delineations. Conversely, the perceptual loss using MedicalNet model, while producing sharper images, tends to introduce artifacts and also falls short in defining precise anatomical structure borders. In contrast, the TotalSegmentator model delivers excellent visual quality, featuring sharp reconstruction and precise delineation of structures, notably in the colon region as depicted in the zoomed-in preview.

\begin{table*}[!h]
\centering
\caption{Comparison of 3D nnU-Net models with different losses for pelvic synthesis using Dice score and NSD on TotalSegmentator segmentations between real CT and synthesized CT.}
\begin{footnotesize}
\begin{tabular}{l|c|c|c|c}
\textbf{Loss} & \multicolumn{2}{c|}{\textbf{Bones}} & \multicolumn{2}{c}{\textbf{Muscles}} \\ \hline
 & \textbf{Dice} & \textbf{NSD} & \textbf{Dice} & \textbf{NSD} \\ \hline
L1 & \cellcolor{3} $0.738 \pm 0.239$ & \cellcolor{3} $0.774 \pm 0.222$ & \cellcolor{3} $0.687 \pm 0.210$ & \cellcolor{3} $0.719 \pm 0.208$ \\
Perceptual & \cellcolor{4} $0.727 \pm 0.241$ & \cellcolor{4} $0.759 \pm 0.236$ & \cellcolor{4} $0.646 \pm 0.237$ & \cellcolor{4} $0.698 \pm 0.225$ \\
AFP TotalSeg & \cellcolor{1} $\mathbf{0.780 \pm 0.203}$ & \cellcolor{1} $\mathbf{0.823 \pm 0.200}$ & \cellcolor{1}$0.724 \pm 0.193$ & \cellcolor{1}$0.762 \pm 0.188$ \\
L1 + AFP TotalSeg & \cellcolor{1} $0.772 \pm 0.194$ & \cellcolor{1} $0.817 \pm 0.196$ & \cellcolor{1} $\mathbf{0.728 \pm 0.195}$ & \cellcolor{1} $\mathbf{0.768 \pm 0.185}$
\end{tabular}
\end{footnotesize}
\label{tab:totalseg_pelvis}
\end{table*}

Table \ref{tab:totalseg_pelvis} and bottom row of Fig. \ref{fig:pelvis} provide a comparative evaluation of the models on anatomical regions using the TotalSegmentator pipeline with Dice score and NSD. Aligning with the visual analysis, the model trained with the AFP loss using TotalSeg's embeddings provide the best performances in every region. Combining the AFP loss with a L1 loss does not significantly affect the reconstruction performance of anatomical structures. The model trained with MedicalNet perceptual loss provides poor results overall, while the model trained with L1 loss delivers average performance in each region.

\section{Discussion}
\subsection{Feature-Prioritized Loss: A Simple Yet Effective Approach}
This study presents an exploration into medical image synthesis with a focus on MR to CT translation for both lung and pelvic regions, specifically aiming to enhance localized structural detail reconstruction through a novel anatomical feature-prioritized loss (AFP loss). Our findings reveal that this method effectively reconstructs targeted anatomical features and can either complement or replace global reconstruction strategies such as L1 loss or perceptual loss. The AFP loss, while conceptually simple in its implementation—essentially applying L1 loss in the embeddings of a pre-trained segmentation network—demonstrates a significant impact on the representation of anatomical structures. This straightforward approach allows the synthesis model to develop a more advanced understanding and representation of complex medical structures, highlighting the potential of leveraging domain-specific knowledge in synthesis networks. 

\subsection{Comparison with Traditional Methods and Metrics}
Global reconstruction loss is commonly expressed using a perceptual loss, an L1 loss, or an L2 loss. The standard L1 loss, directly applied to voxel intensities, is usually defined as the primary global reconstruction loss, as it is more robust to outliers in contrast to the L2 loss, which also tends to yield blurry results \cite{pix2pix}. As evident in Table \ref{tab:mae_airways_lungs} and Table \ref{tab:mae_pelvis}, standalone L1 loss usually outperform our proposed AFP loss in terms of global intensity-based metrics. This is because L1 and L2 losses are inherently designed to optimize pixel-wise intensity minimization, directly influencing global metrics such as Mean Absolute Error (MAE) and Structural Similarity Index (SSIM). However, this optimization often comes at the cost of high-frequency details and fine structural preservation, as these losses tend to average plausible intensities. In contrast, our AFP loss does not explicitly minimize pixel intensities but prioritizes the preservation of structural information. While the AFP loss may not always achieve the absolute lowest MAE or highest SSIM, it excels at reconstructing fine structures, which are crucial for downstream tasks in medical imaging. Ultimately, we argue that the choice of loss function should be guided by the specific requirements of the application. In our context, ensuring the accurate reconstruction of anatomical details is paramount, even if it means a marginal difference in global intensity-based metrics.
On the other hand, perceptual losses are often more robust and deliver better results than L1 loss but are usually limited to 2D applications \cite{longuefosse_perceptual}), unlike our AFP loss, which can easily be used in 3D using trained segmentation networks. While some recent works in medical imaging have attempted to define task-aware or anatomy-aware losses, they often rely on the final output segmentation masks, discarding the internal multi-scale representations learned by the network. Our AFP loss is, to our knowledge, the first to explicitly leverage the intermediate features of pre-trained segmentation networks to guide synthesis, drawing richer anatomical priors from the full depth of the encoder-decoder pipeline.

For global image synthesis quality, traditional metrics such as MAE and SSIM were employed. However, these intensity-based metrics alone often do not adequately capture the nuanced needs of medical imaging, such as the accurate delineation of boundaries between different tissues or the identification of lesions. This underscores the necessity of using structure-specific metrics such as Dice score and Normalized Surface Distance (NSD) for evaluating the quality of local reconstructions, particularly focusing on preserving anatomical structural details that are vital for clinical analysis. 

\subsection{Performance and Insights}
Our experiments indicate that the AFP loss excels in refining the reconstruction of targeted structures, regardless of the dataset (lung or pelvis) or the method used (GAN or U-Net). In the lung region, we evaluated our models on three segmentation tasks: airways and sain bronchi using the NaviAirway pipeline, airway lesions with the Holistic Airway Lesions (HAL) pipeline, and larger anatomical regions (muscles, organs, bones) using the TotalSegmentator pipeline. As expected, the best model for each task is often the one with embeddings closely aligned with the task being evaluated. For instance, for muscle and organ reconstruction based on TotalSegmentator segmentation, the most effective model utilizes AFP loss with TotalSegmentator's embeddings.

However, surprisingly, some networks with non-specialized embeddings can produce high-quality reconstructions or even outperform specialized models when combined with them. For example, the combination of NaviAirway and HAL's embeddings achieved the best performance for airway reconstruction, surpassing the AFP loss with NaviAirway's embeddings alone. Additionally, for bone reconstruction, AFP loss combining NaviAirway and HAL's embeddings yielded excellent results, despite these two networks originally being specialized in fine structures within the lungs, which are quite different from bone segmentation. These findings highlight the advanced understanding and representation capabilities of pre-trained segmentation networks and underscore the value of using them as replacements for perceptual loss or L1 loss.

Moreover, while a general model like TotalSegmentator is not perfectly suited for fine structure reconstruction, it serves as an excellent starting point for coarser tasks. It systematically outperforms L1 loss and MedicalNet perceptual loss in every segmentation task, and provides sharper and more realistic visual reconstructions compared to these other models.

In the pelvic region, models using AFP loss on TotalSegmentator's embeddings and AFP loss combined with L1 loss performed similarly in the downstream task and had similar visual reconstructions, despite having very different results in the intensity-based evaluation. This discrepancy once again calls into question the relevance of these metrics, as they do not accurately reflect the medical and anatomical accuracy of the synthesized volumes.

Initially, perceptual loss, typically derived from pre-trained 2D networks like VGG, was considered. However, when applied using a 3D MedicalNet model specifically adapted for medical data, the results were underwhelming, even when compared to the traditional L1 loss. This suggests ongoing challenges in using medical image datasets for perceptual loss due to their limited diversity and smaller scale compared to extensive, varied datasets like ImageNet \cite{dataset_medical}. Consequently, we recommand relying on the L1 loss or AFP loss with TotalSegmentator's embeddings for coarse reconstructions, which consistently delivered superior image quality.

Moreover, the elimination of the discriminator, particularly when using nnU-Net as a generator only, resulted in even better outcomes. This highlights the potential need to reconsider traditional cost functions in medical synthesis. Instead of focusing on output likelihood — which is typical with discriminators — a more clinically relevant approach would focus on preserving anatomical structures and ensuring low false-positive rates. This aligns directly with the clinical objective of synthesizing volumes that not only possess high signal quality but also maintain the integrity of anatomical details, facilitating advanced analysis by radiologists.

From a clinical perspective, our approach has the potential to significantly improve diagnostic workflows by providing more anatomically faithful synthetic images. For instance, in cystic fibrosis patients, the precise delineation of bronchiectasis from MRI scans is essential for assessing disease severity and monitoring progression \cite{cystic}. Traditionally, CT scans are used for this purpose due to their superior spatial resolution, but they expose patients to ionizing radiation. By enabling MR-to-CT synthesis with enhanced structural accuracy, particularly in fine airway reconstructions, our method can help mitigate the need for repeated CT acquisitions, offering a safer, non-invasive alternative for longitudinal disease monitoring.

\subsection{Dependency on Segmentation Networks}
The effectiveness of the feature-prioritized (AFP) loss is closely linked to the performance of the pre-trained segmentation models it relies on. Since the AFP loss depends on the gradients provided by these models, its success in enhancing the synthesis task hinges on the quality of the segmentation network's output. Our experiments demonstrated that integrating AFtP loss into synthesis networks, including both GAN-based models (e.g., pix2pixHD, SPADE) and CNN-based models (e.g., adapted nnU-Net), significantly improves the reconstruction of specific anatomical structures. However, the reliance on pre-trained segmentation models introduces a limitation: the AFP loss may underperform in situations where such models are unavailable or do not perform well on the target anatomy. This is particularly relevant in clinical scenarios where customized segmentation models may not be readily accessible or where the available models are suboptimal. To address this, we note that publicly available, general-purpose segmentation networks, such as TotalSegmentator, offer a potential solution. These networks, trained on a broad range of anatomical regions, can serve as effective feature extractors. In cases where fine-tuned models for specific regions or pathologies are not available, combining multiple AFP losses tailored to different clinical scenarios could further enhance the synthesis process. For example, one might apply the NaviAirway model for fine structure preservation in the lungs while using TotalSegmentator for coarser anatomical regions outside the lungs. This hybrid approach could offer a more targeted, clinically relevant solution, which we outline as a promising direction for future research.

\section{Conclusion}
This work introduces a novel Anatomical Feature-Prioritized (AFP) loss, which leverages embeddings from pre-trained segmentation networks to guide MR-to-CT synthesis toward more accurate reconstruction of anatomical details. Our results across lung and pelvic datasets demonstrate that AFP loss improves the preservation of fine and coarse structures compared to traditional intensity-based and perceptual losses. While the approach depends on the availability of effective segmentation models, publicly available generalist networks can serve as useful feature extractors, and combining embeddings from multiple models shows promise for diverse anatomical regions. This method offers a straightforward yet effective way to incorporate domain-specific anatomical knowledge into synthesis networks, helping to better meet clinical requirements. Future research will focus on extending AFP loss to additional segmentation models and exploring dynamic weighting schemes for multi-scale features to further enhance synthesis quality.

\section*{Acknowledgments}
This work was granted access to the HPC resources of IDRIS under the allocation 2022-AD011013848R1 made by GENCI.

\section*{References}
\bibliographystyle{dcu}
\bibliography{ref}

\end{document}